\documentclass[twocolumn,pra,superscriptaddress]{revtex4-2}
\usepackage{graphicx,subfigure}  
\usepackage{epstopdf}
\usepackage{amsfonts,amsmath,amssymb, amsthm}    
\usepackage{dsfont}                      
\usepackage[dvipsnames]{xcolor}          
\usepackage[
colorlinks=true,
linkcolor=blue,
citecolor=blue,
urlcolor=blue]{hyperref}                
\usepackage{comment}                    
\usepackage{physics}                    
\usepackage{bm}
\usepackage{mdframed}
\usepackage{enumitem}
\usepackage{soul}

\newcommand{\js}[1]{\textcolor{blue}{[JS: #1]}} 
\newcommand{\kb}[1]{\textcolor{teal}{[KB: #1]}} 
\newcommand{\chg}[1]{\textcolor{purple}{#1}} 

\newtheorem{theorem}{Theorem}
\newtheorem{lemma}{Lemma}
\newtheorem{definition}{Remark}

\begin{document}

\title{Robust and parallel semi-device-independent self-testing of distinct quantum systems}

\author{Dhananjoy Ghosh}
\email{d25062@students.iitmandi.ac.in}
\affiliation{Center for Quantum Science and Technology, Indian Institute of Technology, Mandi, Himachal Pradesh 175075, India}


\author{Jaskaran Singh}
\email{jaskaran@iitmandi.ac.in}
\affiliation{Center for Quantum Science and Technology, Indian Institute of Technology, Mandi, Himachal Pradesh 175075, India}


\begin{abstract}
Robust certification of multiple quantum systems simultaneously, known as parallel self-testing, has so far been formulated for entangled states certified through Bell inequalities in a device-independent manner, where the systems certified in parallel are copies of a single entangled state. We develop a general framework that robustly and simultaneously certifies multiple indivisible quantum systems that can be distinct from one another, through the violation of non-contextual (NC) inequalities, in a semi-device-independent setting. We prove two results, each with explicit
robustness bounds. First, if individual, and possibly distinct, quantum systems each admit a robust self-test via an NC inequality, then they also admit a robust parallel self-test. Second, we establish a converse result as well. Whenever a composite system that decomposes into smaller (and possibly distinct) subsystems admits a robust self-test, then every subsystem itself admits a robust self-test. We illustrate our framework with an example of a qutrit and a ququart system certified in parallel with each other, among other examples. 
\end{abstract}


\maketitle


\textit{Introduction}.--- 
Certification of quantum states and processes~\cite{S96,EW17,KLSBT23,JS21,CJSA19,PSSA24,AMP12,P26,DNJ26} is an essential ingredient in the development of quantum technologies, where the construction of reliable quantum systems is necessary. Among the available approaches, self-testing~\cite{MY04,SB20,K17,SOA26} achieves a robust certification of pure quantum states and projective measurements based only on the input–output statistics of the devices in use~\cite{APPC21,RP23,CGS17}. It relies on the construction of a witness whose extremal value uniquely identifies the underlying state and measurements, up to local isometries. Different witnesses can certify different quantum systems. Bell inequalities~\cite{Bell64} certify maximally and partially entangled states~\cite{MY04,M12,SASA16,ZCPG19,LWHZ22,JC24,MPW25,K16,BZZXHT25,AMSMSM22}, while non-contextual (NC)~\cite{S05,SDASA26} and Leggett–Garg inequalities~\cite{LG85} certify single quantum systems, in each case together with the associated projective measurements~\cite{KBMA19,PSP24,JFBC24,SSA20,HK23,ASCA21}.  

These witnesses also differ in the assumptions they require. Bell inequality based self-testing is device-independent~\cite{MY04,SB20,KBMZ22}. It requires two spatially separated, non-signaling parties but proceeds under minimal assumptions on the underlying quantum systems~\cite{BCPSW14}. Self-testing from NC inequalities relaxes the requirement of separated parties and hence applies to single, indivisible systems, but at the cost of further assumptions on the measurements, namely compatibility (joint measurability) of the prescribed contexts, repeatability, memoryless-ness and their projective nature~\cite{KBMA19}. It is therefore semi-device-independent. Despite these assumptions, NC-based self-testing has recently been demonstrated experimentally~\cite{HK23} and has been used for semi-device-independent randomness expansion~\cite{JFBC24}, paralleling the role Bell inequalities based self-testing plays in the device-independent setting~\cite{BSCA18,RP23,WBDMS16}.

To achieve scalability, quantum technologies are moving toward architectures comprising several quantum systems, which complicates the problem of certification~\cite{SB20,ABB18,P18}. Standard tomographic schemes require a number of measurements and copies of the quantum state that grows exponentially with the dimension of the underlying Hilbert space~\cite{BCG13,PMSLP24,MPS03}. For such situations, self-testing offers a promising alternative by certifying several systems simultaneously, or in parallel with each other~\cite{SB20}. Frameworks for parallel self-testing of maximally and partially entangled systems have been developed from various Bell inequalities~\cite{RFU13,C17,M17,WBDMS16,M16,BKM19,CN16,AW22,KM18,CGJV20}.

These frameworks are limited in two respects. First, they certify entangled states via Bell inequalities~\cite{C17,M17,M16,ZCPZ19,WBDMS16}. Parallel self-testing of multiple copies of single systems via NC inequalities remains an open problem. Secondly, the constituents certified in parallel are copies of a single system~\cite{BKM19,C17,AW22}. For example it is possible to self-test
$n$ copies of the singlet state, but not a singlet state in parallel with a qutrit, even though each can be self-tested individually. A general framework for self-testing distinct quantum systems, which may belong to different dimensional Hilbert spaces, in parallel is therefore lacking. Such a framework would substantially strengthen certification techniques for complex, multi-component quantum systems, as demanded by emerging quantum technologies.  

In this Letter, we develop such a framework, based on NC inequalities, in the semi-device-independent setting. We prove two results, both with explicit robustness bounds. First, if a collection of (possibly distinct) quantum systems can each be self-tested individually, then they can be self-tested simultaneously. 
This generalizes existing parallel self-testing results to arbitrary heterogeneous compositions. Whereas earlier results are device-independent and based on Bell inequality violations, our framework operates in the semi-device-independent regime and are based on NC inequality violations. Second, we establish a converse as well. If a composite system that decomposes into several smaller (and possibly distinct) subsystems can be robustly self-tested, then every subsystems can itself be robustly self-tested. To the best of our knowledge, no analogous converse, in which certifiability of a composite implies certifiability of a constituent, has been established for self-testing. We illustrate our framework with several examples. In the first example, two qutrit systems are self-tested in parallel via two KCBS inequalities. In the second example, we consider the simplest heterogeneous setting, where a qutrit and a ququart are self-tested in parallel using the KCBS and Bell–CHSH inequalities, respectively~\cite{KCBS08,CHSH69}. Furthermore, we explore several examples in the Appendix, including two parallel Bell–CHSH scenarios and a composite scenario involving the KCBS inequality and a non-self-testing NC inequality.


\textit{Graph theoretic approach to self-testing from NC-inequalities}.---
A non-contextual inequality is a linear constraint that is satisfied by all NC hidden variable (NCHV) models~\cite{KS67} but is violated by quantum theory (QT)~\cite{KCBS08}. With the help of these features, a variety of practical applications have been demonstrated~\cite{UZZK20,JKA17,CSW14,DJKA23,ZJJAJA23,JS21,JSS21,PSA26}, including self-testing~\cite{KBMA19,PSP24,BSCA18}.

Consider a finite set of projective dichotomic measurements $\{\Pi_i\}_{i=1}^N$, having outcomes $a\in \{0,1\}$. Let $\mathcal{G}=\lbrace \mathcal{V},\mathcal{E}\rbrace$ be the graph of exclusivity of $\{\Pi_i\}_{i=1}^N$, where each element of vertex set $\mathcal{V}=\{v_i\}_{i=1}^N$ represents one element of $\{\Pi_i\}_{i=1}^N$. Two vertices $v_i$ and $v_j$ are adjacent, i.e., $(v_i,v_j)\in\mathcal{E}$, if and only if the corresponding projectors are mutually exclusive, i.e. $\text{tr}\left(\Pi_i \Pi_j\right) = 0$. Hereafter, if two vertices are adjacent, they will be denoted by $v_i \sim v_j$.


Let $\mathcal{G}$ be an exclusivity graph with $N\geq 5$ vertices, then an NC inequality can be defined~\cite{CSW14} as
\begin{equation}
    \beta := \sum_{i=1}^N w_i p(1|i)\leq \beta_{\text{NCHV}},
    \label{eq:NC_inequality_main}
\end{equation}   
where $p(1|i)$ is the probability of obtaining the outcome $1$ when measuring the projector $\Pi_i$ and $w_i$ is the weight associated with the corresponding vertex $v_i$ of the exclusivity graph $\mathcal{G}$. Henceforth, we restrict our attention to uniformly weighted NC inequalities and set $w_i=1$ for all $i$. Here, the NC bound is $\beta_{\text{NCHV}} = \alpha(\mathcal{G})$, where $\alpha(\mathcal{G})$ is the independence number of the exclusivity graph $\mathcal{G}$~\cite{CSW14,L79} and the maximum quantum value of $\beta$ is $\beta_{\text{QT}}=\vartheta(\mathcal{G})$, where $\vartheta(\mathcal{G})$ is the Lov\'asz theta number of $\mathcal{G}$~\cite{CSW14,L79}.

Next, we define an ideal quantum realization $\mathcal{S} = \{\rho,\{\Pi_i\}_{i=1}^N\}$ which achieves the value $\beta = \beta_{\text{QT}}$ by a tuple consisting of an underlying quantum state $\rho$ and the set of $N$ projective measurements $\{\Pi_i\}_{i=1}^N$ which act on it. Furthermore, we can also define an implemented quantum realization $\tilde{\mathcal{S}} = \{\tilde{\rho},\{\tilde{\Pi}_i\}_{i=1}^N\}$, which may achieve the value $\beta\leq \beta_{\text{QT}}$ and may differ from the ideal realization. 

Without loss of generality, we can always choose the quantum state to be pure and the measurements to be projective. Indeed, any mixed state can be represented as a pure state in a higher dimensional Hilbert space, while any positive operator value measure can be realized as a projective measurement on an extended Hilbert space. Hereafter, we contrast between the ideal and implemented quantum strategy with a tilde on the latter.

It has been shown in Ref.~\cite{KBMA19}, that the implemented quantum realizations achieving $\beta = \beta_{\text{QT}} - \epsilon$ are $\sqrt{\epsilon}$-close (in trace norm distance) to the ideal quantum realization, a property known as self-testing from NC inequalities. For an arbitrary value of $\epsilon$, we define this property as follows.

\begin{definition}{(Robust self-testing from NC inequalities~\cite{KBMA19})}\label{Self_testing_for_single_state_measurement}
   Consider $\mathcal{G}$ be an exclusivity graph with $N\geq 5$ vertices, and an ideal quantum realization $\left\{\rho = \ket{v_0}\!\bra{v_0},\{\Pi_i=\ket{v_i}\!\bra{v_i}\}_{i=1}^N\right\}$ such that $\text{tr}(\Pi_i\Pi_j)=0$ for all $v_i\sim v_j$ which achieves $\beta = \beta_{\text{QT}}$, then the NC inequality in Eq.~\eqref{eq:NC_inequality_main} provides a robust self-test for this quantum realization, such that for any implemented quantum realization $\tilde{\mathcal{S}} = \left\{\tilde{\rho} = \ket{\tilde{v}_0}\!\bra{\tilde{v}_0},\{\tilde{\Pi}_i = \ket{\tilde{v}_i}\!\bra{\tilde{v}_i}\}_{i=1}^N\right\}$ that also achieves $\beta = \beta_{\text{QT}}-\epsilon$, there exists an isometry $V$ such that $||V\ket{v_i}\!\bra{v_i}V^\dagger-\ket{\tilde{v_i}}\bra{\tilde{v_i}}||\leq \mathcal{O}(\sqrt{\epsilon}),\quad 0\leq i\leq N$.
\end{definition}  

Henceforth, following Remark~\ref{Self_testing_for_single_state_measurement}, we state that a NC inequality $\beta$ provides a $\left(\epsilon, \frac{1}{2}\right)$-robust self-test for an ideal realization $\mathcal{S}$ if the robustness scales as $\mathcal{O}\left(\sqrt{\epsilon}\right)$.

To prove that the NC inequality in Eq.~\eqref{eq:NC_inequality_main} admits a robust self-test for the realization $\mathcal{S}$, we use the semi-definite programming (SDP) framework of Refs.~\cite{CSW14,KBMA19,AHO97}. In this framework we first define the primal SDP for the quantum maximum $\beta_{\mathrm{QT}}=\vartheta(\mathcal{G})$ as
\begin{equation}
\begin{aligned}
\vartheta(\mathcal{G})&=\max\sum_{i=1}^{N}X_{ii}\\
\text{s.t.}\quad
&X_{00}=1,\qquad X\in\mathcal{S}_+^{1+N},\\
&X_{i0}=X_{ii},\quad 1\le i\le N,\\
&X_{ij}=0,\qquad v_i\sim v_j,
\end{aligned}
\label{eq:SDP_for_composite_main}
\end{equation}
where $X$ is the Gram matrix associated with an arbitrary quantum realization. Second, as shown in Ref.~\cite{KBMA19}, the NC inequality $\beta$ robustly self-tests $\mathcal{S}$ if and only if the linear system of equations
\begin{equation}
M_{00}=0,\quad M_{i0}=M_{ii},\quad
M_{ij}=0 ~\forall~ v_i\sim v_j,\quad MZ^*=0,
\label{eq:linear_equation_for_composite_case}
\end{equation}
admits the unique solution $M=0$, where $Z^*$ is the dual optimal solution of the SDP in Eq.~\eqref{eq:SDP_for_composite_main} and $M$ is a variable square matrix of dimension $N + 1$.



\textit{Self testing from $n$-Parallel NC Inequalities}.---
Following the underlying assumptions for exhibiting violations of NC inequalities, we assume that all measurements are ideal, implying that repeatedly performing the same measurement on the same system results in identical outcomes. We also assume that the system being measured has no memory, and each measurement device is used only once. Furthermore, we also consider that different experimental runs are independent and prepared identically, with an unlimited supply of such devices.

Consider a set of $n$ exclusivity graphs $\lbrace \mathcal{G}_k \rbrace_{k = 1}^{n}$ such that the vertex set of each graph $\mathcal{G}_k$ is defined as $\mathcal{V}_k = \lbrace v_{i_k} \rbrace_{i_k}$ with number of vertices $\abs{\mathcal{V}_k} = N_k$. For each exclusivity graph $\mathcal{G}_k$, each vertex $v_{i_k}$ is associated with a projector $\Pi_{i_k}$ such that two projectors corresponding to adjacent vertices are orthogonal. Corresponding to each graph $\mathcal{G}_k$ a NC inequality can be defined as
\begin{equation}
    \beta_k := \sum_{i_k = 1}^{N_k}p(1|i_k) \leq \beta_{\text{NCHV}}^{(k)},
    \label{eq:NC_each_graph}
\end{equation}
where $\beta_{\text{NCHV}}^{(k)} = \alpha(\mathcal{G}_k)$ is the NC bound corresponding to the exclusivity graph $\mathcal{G}_k$ and the maximum quantum value is $\beta_{\text{QT}}^{(k)} = \vartheta(\mathcal{G}_k)$.

While each NC inequality can be tested individually, it is also possible to perform a joint test by constructing a single composite NC inequality that incorporates all individual inequalities being implemented in parallel to each other. The composite NC inequality can be defined by constructing a composite exclusivity graph from the individual exclusivity graphs $\mathcal{G}_k$.

To construct the composite NC scenario, we allow the different systems to be correlated while assuming that no communication is possible between the individual systems~\cite{M17,C17,WBDMS16,M16,BKM19,CN16,AW22,KM18,CGJV20}. The no-communication assumption can be enforced through the no-signalling condition. Under these two assumptions, Ref.~\cite{AFLS15} shows that the non-orthogonality graph corresponding to the parallel composition of two NC scenarios is given by the strong graph product of the individual non-orthogonality graphs. Specifically, if we consider two non-orthogonality graphs, $\overline{\mathcal{G}}_1$ and $\overline{\mathcal{G}}_2$, corresponding to two distinct NC scenarios, then the non-orthogonality graph corresponding to composite NC scenarios can be written as
\begin{equation}
    \overline{\mathcal{G}}
    =
    \overline{\mathcal{G}}_1
    \boxtimes
    \overline{\mathcal{G}}_2,
\end{equation}
where $\boxtimes$ represents the strong product of graphs.

In our framework since we employ orthogonality graphs or exclusivity graphs, we need to determine the graph product corresponding to the parallel composition of NC scenarios.
We note that
\begin{equation}
\begin{aligned}
    \mathcal{G}
    &=
    \overline{
    \overline{\mathcal{G}}_1
    \boxtimes
    \overline{\mathcal{G}}}_2
     \\
    &=
    \overline{\overline{\mathcal{G}}}_1
    \vee
    \overline{\overline{\mathcal{G}}}_2 \\
    &=
    \mathcal{G}_1
    \vee
    \mathcal{G}_2,
\end{aligned}
\end{equation}
where $\vee$ denotes the OR (disjunctive) product of graphs and in the second equality we have used De Morgan's identity~\cite{H11,ADRSW17}.

Hence, the exclusivity graph corresponding to the parallel NC scenario is given by the OR graph product of the individual exclusivity graphs.

For the OR product of two exclusivity graphs, $\mathcal{G}=\mathcal{G}_1\vee\mathcal{G}_2$, the exclusivity relation in the composite graph $\mathcal{G}$ is defined such that two vertices $\lbrace \left(v_{i_1}, v_{i_2}\right), \left(v_{j_1}, v_{j_2}\right)\rbrace \in \mathcal{V}_1 \times \mathcal{V}_2$ are adjacent whenever either $v_{i_1}\sim v_{j_1}$ or $v_{i_2}\sim v_{j_2}$~\cite{H11}.

Using the OR product, we can define the composite exclusivity graph corresponding to parallel tests of individual NC inequalities associated with graphs $\lbrace \mathcal{G}_k \rbrace_k$ as $\mathcal{G} = \mathop{\bigvee}\limits_{k=1}^n \mathcal{G}_k,$ where the vertex set of $\mathcal{G}$ is given as $\mathcal{V} = \mathcal{V}_1 \times \ldots \times \mathcal{V}_n$ with a total number of vertices $N = \abs{\mathcal{V}} = \prod_{k = 1}^{n} N_k$. To each vertex $v_i = \left(v_{i_1},\ldots, v_{i_n}\right) \in \mathcal{V}$ we associate a projector $\Pi_i = \bigotimes_{k = 1}^{n}\Pi_{i_k}$. Finally, the composite NC inequality corresponding to $n$-parallel NC inequalities can be defined following Eq.~\eqref{eq:NC_inequality_main}.

Corresponding to every NC inequality in the composite NC inequality there exists an ideal quantum realization denoted by $\mathcal{S}_{k} = \lbrace\rho_{k},\{\Pi_{i_k}\}_{i_k=1}^{N_k}\rbrace$, for which the NC inequality $\beta_k = \beta^{(k)}_{\text{QT}}$. As such the composite ideal quantum realization for which the $n$-parallel NC inequality $\beta = \beta_{\text{QT}}$ can be denoted by $\mathcal{S} = \lbrace\rho,\{\Pi_i\}_{i=1}^N\rbrace$ where $\rho = \bigotimes_{k = 1}^{n} \rho_k$ and $\Pi_i = \bigotimes_{k = 1}^{n} \Pi_{i_k}$. Here we note that the quantum state and the projectors in the composite ideal quantum realization are tensor products of the ideal quantum states and projectors of component NC inequalities, respectively. We can also define a composite implemented quantum realization $\tilde{\mathcal{S}} = \lbrace\tilde{\rho},\{\tilde{\Pi}_i\}_{i=1}^N\rbrace$ which also achieves $\beta \leq \beta_{\text{QT}}$ such that the underlying quantum state $\tilde{\rho}$ and the projectors $\tilde{\Pi}_i$ may not admit a tensor product representation in terms of individual quantum states and projectors as the composite ideal quantum realization.

We can now state our main results of self-testing of $n$-parallel NC inequalities. We defer all proofs to the Appendix~\ref{appen:proofs}.

\begin{theorem}\label{thm:1}
     Consider a set of $n$ NC inequalities $\lbrace \beta_k \rbrace_k$ associated with exclusivity graphs $\lbrace \mathcal{G}_k \rbrace_k$ respectively, such that each NC inequality $\beta_k$ gives a $(\epsilon_k,\frac{1}{2})$-robust self-test for an ideal quantum realization $\mathcal{S}_k=\{\rho_k = \ket{v_{0_k}}\!\bra{v_{0_k}},\{\Pi_{i_k} = \ket{v_{i_k}}\!\bra{v_{i_k}}\}_{i_k=1}^{N_k}\}$. Then, one can define a composite NC inequality $\beta$ associated with a composite graph $\mathcal{G} = \bigvee_{k=1}^{n}\mathcal{G}_k$. If the composite ideal quantum realization $\mathcal{S} = \lbrace \rho = \bigotimes_{k = 1}^{n}\ket{v_{0_k}}\!\bra{v_{0_k}}, \lbrace \Pi_i = \bigotimes_{k = 1}^{n}\ket{v_{i_k}}\!\bra{v_{i_k}} \rbrace_{i}\rbrace$ achieves $\beta = \beta_{\text{QT}}$ such that $\beta_{\text{QT}}>\beta_{\text{NCHV}}$, then the NC inequality $\beta$ provides a $(\epsilon,\frac{1}{2})$-robust self-test for the quantum realization $\mathcal{S}$. Explicitly, for every implemented quantum strategy $\tilde{\mathcal{S}} = \left\{\tilde{\rho} = \ket{\tilde{v}_0}\!\bra{\tilde{v}_0},\{\tilde{\Pi}_i = \ket{\tilde{v}_i}\!\bra{\tilde{v}_i}\}_{i}\right\}$ that attains $\beta=\beta_{\text{QT}}-\epsilon$ then, there exists an isometry $V$ such that
    \begin{equation}
\norm{V\ket{v_i}\!\bra{v_i}V^\dagger-\ket{\tilde{v_i}}\bra{\tilde{v_i}}}\leq \mathcal{O}(\sqrt{\epsilon}), \quad 0\leq i\leq N.
        \label{eq:parallel_self_test_robust}
    \end{equation}
\end{theorem}

We emphasize here that the proof of Theorem~\ref{thm:1} is non-trivial. For the OR product of graphs, both the independence number $\alpha$ and the Lov\'asz theta number $\vartheta$ are multiplicative~\cite{ADRSW17}, so the condition $\vartheta(\mathcal{G})>\alpha(\mathcal{G})$ is always valid for Theorem~\ref{thm:1}, but this condition itself is not sufficient for self-testing. There exist NC inequalities that exhibit a quantum violation but nevertheless fail to self-test the underlying quantum realization~\cite{KMAAL19}.

It should be noted that Theorem~\ref{thm:1} is not restricted to parallel repetitions of identical NC inequalities. It also applies to heterogeneous scenarios involving distinct NC inequalities and distinct quantum realizations. In such cases, the theorem provides a framework for certifying a composite realization obtained from different constituent realizations through the OR graph product.

So far, we have only shown that robust self-testing of individual quantum realizations implies parallel self-testing of the composite quantum realization. A natural question is whether the converse is also true? Specifically, if a graph $\mathcal{G}$ admits a self-test, then does it imply that all individual NC inequalities corresponding to $\mathcal{G}_k$, such that $\mathcal{G} = \bigvee_{k = 1}^n \mathcal{G}_k$, also admit a self-test? In the following theorem, we prove that if the composite NC inequality corresponding to $\mathcal{G}$ admits a robust self-test, then it implies that every individual NC inequality corresponding to $\mathcal{G}_k$ also admits a robust self-test.

\begin{theorem}\label{thm:2}
  Consider an NC inequality $\beta$ associated with an exclusivity graph $\mathcal{G}$ that provides a $(\epsilon,\frac{1}{2})$-robust self-test for an ideal quantum realization $\mathcal{S}=\{\rho,\{\Pi_i\}_{i=1}^N\}$. If $\mathcal{G} = \bigvee_{k=1}^{n}\mathcal{G}_k$, then every individual NC inequality $\beta_k$ corresponding to the exclusivity graph $\mathcal{G}_k$ provides a $(\epsilon_k,\frac{1}{2})$-robust self-test for the corresponding quantum realization $\mathcal{S}_k=\{\rho_k,\{\Pi_{i_k}\}_{i_k}\}$, provided that $\mathcal{S}_k$ achieves $\beta_k=\beta_{\text{QT}}^{(k)}$, such that $\beta_{\text{QT}}^{(k)} >\beta_{\text{NCHV}}^{(k)}$. Explicitly, for every implemented quantum strategy $\tilde{S}_k = \{\tilde{\rho}_k,\{\tilde{\Pi}_{i_k}\}_{i_k}\}$ that attains $\beta_k=\beta_{\text{QT}}^{(k)}-\epsilon_k$, there will exist an isometry $V_k$ where
\begin{equation}
\norm{V_k\ket{v_{i_k}}\!\bra{v_{i_k}}V_k^\dagger-\ket{\tilde{v}_{i_k}}\bra{\tilde{v}_{i_k}}}\leq \mathcal{O}(\sqrt{\epsilon_k}), \quad \forall~k,
\end{equation}
where $0\leq i_k\leq N_k$.
\label{eq:individual_self_test_robust}
\end{theorem}



\textit{Applications}.--- Consider a composite scenario consisting of two KCBS inequalities~\cite{CSW14,KCBS08}, where each KCBS inequality can be written as
    $\beta_j=\sum_{i_j=1}^5 p(1|i_j) \leq 2$, $j = 1,2$
where $\beta^{(j)}_{\text{NCHV}}=2$, while $\beta^{(j)}_{\text{QT}}=\sqrt{5}$, which can be achieved by a pure qutrit (three-dimensional) quantum state with the ideal quantum realization given by 
$\mathcal{S}_j=\{ \ket{v_{0_j}}\!\bra{v_{0_j}},\{ \ket{v_{i_j}}\!\bra{v_{i_j}}\}_{i_j=1}^{5}\}$, where $\ket{v_{0_j}}$ and $\lbrace \ket{v_{i_j}}\rbrace_{i_j = 1}^{5}$ are given in the Appendix~\ref{appsubsec:Parallel_KCBS}~\cite{KBMA19}. The exclusivity graph associated with the KCBS inequality is the five-vertex regular cyclic graph, denoted by $C_5$.

\begin{figure}
    \centering
    \includegraphics[width=0.85\linewidth]{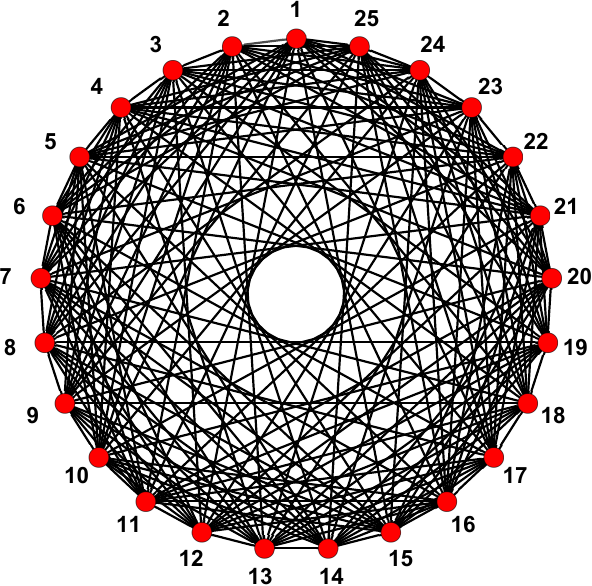}
    \caption{The exclusivity graph $\mathcal{G}$ associated with two parallel KCBS inequalities.}
    \label{fig:ex_graph_2_KCBS}
\end{figure}

Following our framework, corresponding to the two parallel copies of KCBS inequalities, we first construct a composite inequality whose exclusivity graph is obtained as the OR graph product of the exclusivity graphs associated with the individual inequalities, i.e., $\mathcal{G}=C_5\vee C_5$, as shown in Fig.~\ref{fig:ex_graph_2_KCBS}. The resulting graph consists of $N = 25$ vertices and $200$ edges, with $\alpha(\mathcal{G})=4$, and $\vartheta(\mathcal{G})=5$, which can be achieved by the composite qutrit-qutrit realization
    $\mathcal{S}
    =
    \left\{
    \ket{v_0}\!\bra{v_0},
    \left\{\ket{v_i}\!\bra{v_i}\right\}_{i=1}^{25}
    \right\}$,
where $ \ket{v_0}\!\bra{v_0}=
\ket{v_{0_1}}\!\bra{v_{0_1}} \otimes \ket{v_{0_2}}\!\bra{v_{0_2}},$ and $ \ket{v_i}\!\bra{v_i}=\ket{v_{i_1}}\!\bra{v_{i_1}} \otimes \ket{v_{i_2}}\!\bra{v_{i_2}}$. Since $\beta_{\text{QT}}>\beta_{\text{NCHV}}$, the condition of Theorem~\ref{thm:1} is satisfied. Using the exclusivity conditions of $\mathcal{G} = C_5\vee C_5$, we construct the corresponding composite SDP, given in Eq.~\eqref{eq:SDP_for_composite_main}, and obtain a dual optimal solution  
\begin{equation}
    Z^*=\left( \begin{tabular}{c|c}
         $5$ & $-J^{\intercal}$ \\
         \hline
        $-J$ & $z$
    \end{tabular}\right),
    \label{eq:Z25_para_KCBS}
\end{equation}
where $z=\mathds{1}_5\otimes \mathds{1}_5+b_3\left(A_5\otimes \mathds{1}_5+\mathds{1}_5\otimes A_5\right)+b_1\left(A_5\otimes A_5\right)+b_2\left(\bar A_5\otimes A_5+A_5\otimes \bar A_5\right)$ with $
b_1 = c b_3$, $
b_2 = \frac{1}{2}-\frac{b_3}{2c}$, and $c=\frac{5-\sqrt{5}}{2\sqrt{5}}$. The parameter $b_3$ may be chosen arbitrarily, provided that the resulting matrix $Z^*$ remains positive semidefinite. Here, $\mathds{1}_5$ is the $5 \times 5$ identity matrix, $J=(1,\ldots,1)^{\intercal}\in\mathbb{R}^{25}$ denotes the all-ones column vector, while $A_5$ and $\bar A_5$ denote the adjacency matrix and complement adjacency matrix of the graph $C_5$, respectively.


Next, we construct the variable matrix $M$ using Eq.~\eqref{eq:linear_equation_for_composite_case} together with the exclusivity constraints of $\mathcal{G}=C_5\vee C_5$. To solve the linear system $MZ^*=0$, we first reformulate it as a homogeneous system $Ax=0$, where $A$ is a numerical matrix and $x$ is the column vector containing all free variables of $M$. While in this particular case the homogeneous system $Ax=0$ can be solved exactly, this problem becomes intractable for increasingly complex graphs. In such situations, finding a closed form solution of the dual SDP is no longer possible. As a result, we have to work around with the numerical solution, which in general will not give exact solutions to $Ax=0$. For that purpose, we determine the rank of $A$ using singular value decomposition. Singular values equal to zero indicate that the matrix $A$ is rank-deficient and, consequently, admits nontrivial solutions to $Ax=0$. However, as mentioned since a closed form expression for the dual optimal solution is not available, we consider a threshold value to identify (near) zero singular values. This threshold value depends on the numerical precision of the optimization software (in our case MOSEK~\cite{mosek25}). In our case, depending on the precision, we consider the threshold to be $10^{-8}$ and treat all singular values smaller than this as zero. Conversely, singular values larger than this threshold indicate that $A$ is numerically full rank, implying that the trivial solution $x=0$ is unique. Further details of this procedure can be found in our uploaded documentation of the code in Ref.~\cite{GBS26}. Solving the resulting system of equations, we find that all singular values are significantly larger than the threshold value (see Ref.~\cite{GBS26}) which indicates that the only admissible solution is the trivial one $M=0$. This further indicates that the primal optimal solution of the SDP corresponding to the composite two parallel KCBS scenario is unique. Therefore, the composite inequality also self-tests the composite qutrit-qutrit realization $\mathcal{S}$. Moreover, any other implemented quantum realization $\tilde{\mathcal{S}}$ achieves $\beta=\beta_{\text{QT}}-\epsilon$, then there exists an isometry $V$, such that
\begin{equation}
    \norm{V\ket{v_i}\!\bra{v_i}V^\dagger-\ket{\tilde{v_i}}\bra{\tilde{v_i}}}\leq \mathcal{O}(\sqrt{\epsilon}), \quad 0\leq i\leq 25.
\end{equation}

Here, it should be noted that we consider that the two KCBS experiments are performed in parallel within the same laboratory and the equivalence between the ideal and implemented realizations is certified by a single global isometry $V$ such that $V:\mathcal{H}\rightarrow \tilde{\mathcal{H}}$, where $\mathcal{H}$ is the joint Hilbert space of the composite qutrit-qutrit realization and $\tilde{\mathcal{H}}$ is the Hilbert space for the implemented realization.

Consider another composite scenario consisting of one KCBS and one Bell-CHSH inequality, where the KCBS inequality is as described above while the Bell-CHSH inequality~\cite{CSW14,Bell64,CHSH69} can be expressed as $\beta_2=\sum_{i_2=1}^8p(1|i_2)\leq 3,$
where, $\beta^{(2)}_{\text{NCHV}}=3$ and $\beta^{(2)}_{\text{QT}}=2+\sqrt{2}$ which can be achieved by a ququart (four-dimensional) quantum state and an ideal quantum realization $\mathcal{S}_2=\{ \ket{v_{0_2}}\!\bra{v_{0_2}},\{ \ket{v_{i_2}}\!\bra{v_{i_2}}\}_{i_2=1}^{8}\}$, where the explicit form of the vectors $\ket{v_{0_2}}$ and $\lbrace \ket{v_{i_2}}\rbrace_{i_2 = 1}^{8}$ are given in the Appendix~\ref{appsubsec:Parallel_CHSH}.
The exclusivity graph associated with the Bell-CHSH inequality is the eight-vertex circulant graph denoted by $C_{i_8}(1,4)$. 

Individually, the KCBS inequality self-tests the ideal realization $\mathcal{S}_1$, while the Bell-CHSH inequality self-tests the ideal realization $\mathcal{S}_2$.

To show parallel self-testing, we first construct a composite NC inequality whose exclusivity graph is given by the OR graph product of the exclusivity graphs corresponding to the KCBS and Bell-CHSH inequalities, i.e., $\mathcal{G}=C_5 \vee C_{i_8}(1,4)$. The resulting graph contains $N=40$ vertices and $500$ edges, with $\alpha(\mathcal{G}) = 6$ and $\vartheta(\mathcal{G}) = \sqrt{5}(2+\sqrt{2}) \approx 7.63$, which can be achieved by the composite qutrit-ququart realization
    $\mathcal{S}
    =
    \left\{
    \ket{v_0}\!\bra{v_0},
    \left\{\ket{v_i}\!\bra{v_i}\right\}_{i=1}^{40}
    \right\}$,
where $ \ket{v_0}\!\bra{v_0}=
\bigotimes_{k=1}^{2}
\ket{v_{0_k}}\!\bra{v_{0_k}},$ and $ \ket{v_i}\!\bra{v_i}=\bigotimes_{k=1}^{2}\ket{v_{i_k}}\!\bra{v_{i_k}}$. Since $\beta_{\text{QT}}>\beta_{\text{NCHV}}$, the condition of Theorem~\ref{thm:1} is satisfied. Using the exclusivity conditions of $\mathcal{G}=C_5 \vee C_{i_8}(1,4)$, we construct the corresponding composite SDP, given in Eq.~\eqref{eq:SDP_for_composite_main}, and numerically determine an optimal dual solution $Z^*$.

Next, we construct the variable matrix $M$ using Eq.~\eqref{eq:linear_equation_for_composite_case} together with the exclusivity constraints of $\mathcal{G}=C_{5}\vee C_{i_8}(1,4)$. We then solve the linear system $MZ^*=0$. As before, we reformulate the problem as a homogeneous linear system $Ax=0$ and analyze the corresponding coefficient matrix $A$. Since the dual optimal solution is obtained numerically from the SDP solver, $A$ is also numerical. Accordingly, we determine its rank via singular value decomposition. In the present case, we find that all singular values are significantly larger than the threshold value, which indicates that $A$ is numerically full rank~\cite{GBS26}, and hence the only admissible solution is $M=0$, which implies that the primal optimal solution of the SDP corresponding to the composite KCBS and Bell-CHSH scenario is unique. It then follows from Theorem~\ref{thm:1} that the associated composite inequality self-tests the composite qutrit-ququart realization $\mathcal{S}$ with robustness $\mathcal{O}(\sqrt{\epsilon})$.

Here, as in the previous example, the equivalence between the ideal and implemented realizations is certified by a global isometry $V:\mathcal{H}\rightarrow\tilde{\mathcal{H}}$, acting on the entire composite Hilbert space. This is particularly relevant in the present case, where the Bell-CHSH scenario is realized by a ququart (four-dimensional) rather than by two qubits. Consequently, the composite NC inequality self-tests the composite qutrit-ququart realization globally, without assuming any internal tensor-product structure for the ququart subsystem.

We explore further examples in the Appendix. In Appendix~\ref{appsubsec:Parallel_CHSH}, we show parallel self-testing of multiple ququarts using the Bell-CHSH inequality. Furthermore, in Appendix~\ref{appsubsec:Parallel_KCBS_C6}, we also explore (non-)self-testing of qutrits via the KCBS inequality in parallel with a non-self-testing NC inequality. We show that this composite scenario does not provide a self-test. This example further supports our results that all individual NC inequalities must exhibit self-testing for the composite to exhibit a self-test.


\textit{Conclusion}.--- 
In this work, we introduced a general framework for parallel self-testing of distinct quantum systems within the graph-theoretic approach of Ref.~\cite{KBMA19}. We showed that whenever a set of individual NC inequalities each admit a robust self-test, the composite NC inequality obtained by composing them in parallel also admits a robust self-test, certifying the tensor product of the individual realizations. Crucially, the constituent systems need neither be identical nor entangled, in contrast to existing parallel self-testing results, which are restricted to multiple copies of a single entangled state. We further established a converse result as well. If a composite NC inequality robustly self-tests a composite realization, then each constituent inequality robustly self-tests its underlying realization. To the best of our knowledge, this is the first such result relating the certifiability of a composite system to that of its parts. We illustrated our results by first considering an example of two qutrits, self-tested in parallel from two KCBS inequalities, and second considering a heterogeneous scenario consisting of a qutrit and a ququart, self-tested in parallel from the KCBS and Bell-CHSH inequalities. Further examples are provided in the Appendix.

Our results open several directions. The most immediate is randomness expansion. Bell inequalities based self-testing underlies device-independent randomness expansion, and NC inequality based self-testing has been shown to certify randomness in the semi-device-independent setting~\cite{JFBC24}. Our framework can lift this to the parallel regime. We expect this to yield parallel semi-device-independent randomness expansion protocols whose certified randomness scales with the composite realization. Quantifying the achievable rates in such scenarios is a natural next step.

Beyond randomness, the framework is equally applicable to the certification of scalable quantum hardware. As quantum processors integrate many distinct components, certifying them jointly, rather than tomographically and one at a time, becomes essential. Our result shows that joint certification of distinct subsystems reduces to certifying them individually. A failure to certify any constituent system obstructs certification of the whole, which suggests its use as a diagnostic for localizing faults in complex devices.

\textit{Acknowledgments}.--- 
The authors sincerely thank Kishor Bharti for his valuable discussions and suggestions on this manuscript. D.G. acknowledges Apurav Tehri for several helpful discussions. D.G. also acknowledges financial support from the IIT Mandi HTRA fellowship. J.S. acknowledges financial support from the IIT Mandi seed grant project No. IITM/SG/JSN/168.

\begin{thebibliography}{76}%
\makeatletter
\providecommand \@ifxundefined [1]{%
 \@ifx{#1\undefined}
}%
\providecommand \@ifnum [1]{%
 \ifnum #1\expandafter \@firstoftwo
 \else \expandafter \@secondoftwo
 \fi
}%
\providecommand \@ifx [1]{%
 \ifx #1\expandafter \@firstoftwo
 \else \expandafter \@secondoftwo
 \fi
}%
\providecommand \natexlab [1]{#1}%
\providecommand \enquote  [1]{``#1''}%
\providecommand \bibnamefont  [1]{#1}%
\providecommand \bibfnamefont [1]{#1}%
\providecommand \citenamefont [1]{#1}%
\providecommand \href@noop [0]{\@secondoftwo}%
\providecommand \href [0]{\begingroup \@sanitize@url \@href}%
\providecommand \@href[1]{\@@startlink{#1}\@@href}%
\providecommand \@@href[1]{\endgroup#1\@@endlink}%
\providecommand \@sanitize@url [0]{\catcode `\\12\catcode `\$12\catcode `\&12\catcode `\#12\catcode `\^12\catcode `\_12\catcode `\%12\relax}%
\providecommand \@@startlink[1]{}%
\providecommand \@@endlink[0]{}%
\providecommand \url  [0]{\begingroup\@sanitize@url \@url }%
\providecommand \@url [1]{\endgroup\@href {#1}{\urlprefix }}%
\providecommand \urlprefix  [0]{URL }%
\providecommand \Eprint [0]{\href }%
\providecommand \doibase [0]{https://doi.org/}%
\providecommand \selectlanguage [0]{\@gobble}%
\providecommand \bibinfo  [0]{\@secondoftwo}%
\providecommand \bibfield  [0]{\@secondoftwo}%
\providecommand \translation [1]{[#1]}%
\providecommand \BibitemOpen [0]{}%
\providecommand \bibitemStop [0]{}%
\providecommand \bibitemNoStop [0]{.\EOS\space}%
\providecommand \EOS [0]{\spacefactor3000\relax}%
\providecommand \BibitemShut  [1]{\csname bibitem#1\endcsname}%
\let\auto@bib@innerbib\@empty
\bibitem [{\citenamefont {Spiller}(1996)}]{S96}%
  \BibitemOpen
  \bibfield  {author} {\bibinfo {author} {\bibfnamefont {T.}~\bibnamefont {Spiller}},\ }\bibfield  {title} {\bibinfo {title} {Quantum information processing: cryptography, computation, and teleportation},\ }\href {https://doi.org/10.1109/5.546399} {\bibfield  {journal} {\bibinfo  {journal} {Proc. IEEE}\ }\textbf {\bibinfo {volume} {84}},\ \bibinfo {pages} {1719} (\bibinfo {year} {1996})}\BibitemShut {NoStop}%
\bibitem [{\citenamefont {Epstein}\ and\ \citenamefont {Whaley}(2017)}]{EW17}%
  \BibitemOpen
  \bibfield  {author} {\bibinfo {author} {\bibfnamefont {J.~M.}\ \bibnamefont {Epstein}}\ and\ \bibinfo {author} {\bibfnamefont {K.~B.}\ \bibnamefont {Whaley}},\ }\bibfield  {title} {\bibinfo {title} {Quantum speed limits for quantum-information-processing tasks},\ }\href {https://doi.org/10.1103/PhysRevA.95.042314} {\bibfield  {journal} {\bibinfo  {journal} {Phys. Rev. A}\ }\textbf {\bibinfo {volume} {95}},\ \bibinfo {pages} {042314} (\bibinfo {year} {2017})}\BibitemShut {NoStop}%
\bibitem [{\citenamefont {Knörzer}\ \emph {et~al.}(2026)\citenamefont {Knörzer}, \citenamefont {Liu}, \citenamefont {Schiffer},\ and\ \citenamefont {Tura~Brugués}}]{KLSBT23}%
  \BibitemOpen
  \bibfield  {author} {\bibinfo {author} {\bibfnamefont {J.}~\bibnamefont {Knörzer}}, \bibinfo {author} {\bibfnamefont {X.}~\bibnamefont {Liu}}, \bibinfo {author} {\bibfnamefont {B.}~\bibnamefont {Schiffer}},\ and\ \bibinfo {author} {\bibfnamefont {J.}~\bibnamefont {Tura~Brugués}},\ }\bibfield  {title} {\bibinfo {title} {Distributed quantum information processing: a review of recent progress},\ }\href {http://iopscience.iop.org/article/10.1088/1361-6633/ae74e0} {\bibfield  {journal} {\bibinfo  {journal} {Rep. Prog. Phys.}\ } (\bibinfo {year} {2026})}\BibitemShut {NoStop}%
\bibitem [{\citenamefont {Singh}\ and\ \citenamefont {Bose}(2021)}]{JS21}%
  \BibitemOpen
  \bibfield  {author} {\bibinfo {author} {\bibfnamefont {J.}~\bibnamefont {Singh}}\ and\ \bibinfo {author} {\bibfnamefont {S.}~\bibnamefont {Bose}},\ }\bibfield  {title} {\bibinfo {title} {Non-{G}aussian operations in measurement-device-independent quantum key distribution},\ }\href {https://doi.org/10.1103/PhysRevA.104.052605} {\bibfield  {journal} {\bibinfo  {journal} {Phys. Rev. A}\ }\textbf {\bibinfo {volume} {104}},\ \bibinfo {pages} {052605} (\bibinfo {year} {2021})}\BibitemShut {NoStop}%
\bibitem [{\citenamefont {Kumar}\ \emph {et~al.}(2019)\citenamefont {Kumar}, \citenamefont {Singh}, \citenamefont {Bose},\ and\ \citenamefont {Arvind}}]{CJSA19}%
  \BibitemOpen
  \bibfield  {author} {\bibinfo {author} {\bibfnamefont {C.}~\bibnamefont {Kumar}}, \bibinfo {author} {\bibfnamefont {J.}~\bibnamefont {Singh}}, \bibinfo {author} {\bibfnamefont {S.}~\bibnamefont {Bose}},\ and\ \bibinfo {author} {\bibnamefont {Arvind}},\ }\bibfield  {title} {\bibinfo {title} {Coherence-assisted non-{G}aussian measurement-device-independent quantum key distribution},\ }\href {https://doi.org/10.1103/PhysRevA.100.052329} {\bibfield  {journal} {\bibinfo  {journal} {Phys. Rev. A}\ }\textbf {\bibinfo {volume} {100}},\ \bibinfo {pages} {052329} (\bibinfo {year} {2019})}\BibitemShut {NoStop}%
\bibitem [{\citenamefont {Roy}\ \emph {et~al.}(2024)\citenamefont {Roy}, \citenamefont {Bera}, \citenamefont {Gupta},\ and\ \citenamefont {Majumdar}}]{PSSA24}%
  \BibitemOpen
  \bibfield  {author} {\bibinfo {author} {\bibfnamefont {P.}~\bibnamefont {Roy}}, \bibinfo {author} {\bibfnamefont {S.}~\bibnamefont {Bera}}, \bibinfo {author} {\bibfnamefont {S.}~\bibnamefont {Gupta}},\ and\ \bibinfo {author} {\bibfnamefont {A.~S.}\ \bibnamefont {Majumdar}},\ }\bibfield  {title} {\bibinfo {title} {Device-independent quantum secure direct communication under non-{M}arkovian quantum channels},\ }\href {https://doi.org/10.1007/s11128-024-04397-8} {\bibfield  {journal} {\bibinfo  {journal} {Quantum Inf. Process.}\ }\textbf {\bibinfo {volume} {23}},\ \bibinfo {pages} {170} (\bibinfo {year} {2024})}\BibitemShut {NoStop}%
\bibitem [{\citenamefont {Ac\'{\i}n}\ \emph {et~al.}(2012)\citenamefont {Ac\'{\i}n}, \citenamefont {Massar},\ and\ \citenamefont {Pironio}}]{AMP12}%
  \BibitemOpen
  \bibfield  {author} {\bibinfo {author} {\bibfnamefont {A.}~\bibnamefont {Ac\'{\i}n}}, \bibinfo {author} {\bibfnamefont {S.}~\bibnamefont {Massar}},\ and\ \bibinfo {author} {\bibfnamefont {S.}~\bibnamefont {Pironio}},\ }\bibfield  {title} {\bibinfo {title} {Randomness versus nonlocality and entanglement},\ }\href {https://doi.org/10.1103/PhysRevLett.108.100402} {\bibfield  {journal} {\bibinfo  {journal} {Phys. Rev. Lett.}\ }\textbf {\bibinfo {volume} {108}},\ \bibinfo {pages} {100402} (\bibinfo {year} {2012})}\BibitemShut {NoStop}%
\bibitem [{\citenamefont {Singh}(2026)}]{P26}%
  \BibitemOpen
  \bibfield  {author} {\bibinfo {author} {\bibfnamefont {P.}~\bibnamefont {Singh}},\ }\bibfield  {title} {\bibinfo {title} {Coherence-gated quantum devices via real-time weak measurement},\ }\bibfield  {journal} {\bibinfo  {journal} {arXiv}\ }\href {https://doi.org/10.48550/arXiv.2604.18662} {10.48550/arXiv.2604.18662} (\bibinfo {year} {2026}),\ \Eprint {https://arxiv.org/abs/2604.18662} {arXiv:2604.18662 [quant-ph]} \BibitemShut {NoStop}%
\bibitem [{\citenamefont {Ghosh}\ \emph {et~al.}(2026)\citenamefont {Ghosh}, \citenamefont {J},\ and\ \citenamefont {Singh}}]{DNJ26}%
  \BibitemOpen
  \bibfield  {author} {\bibinfo {author} {\bibfnamefont {D.}~\bibnamefont {Ghosh}}, \bibinfo {author} {\bibfnamefont {N.}~\bibnamefont {J}},\ and\ \bibinfo {author} {\bibfnamefont {J.}~\bibnamefont {Singh}},\ }\bibfield  {title} {\bibinfo {title} {Local finite rate semi device independent randomness expansion from quantum contextuality},\ }\bibfield  {journal} {\bibinfo  {journal} {arXiv}\ }\href {https://doi.org/10.48550/arXiv.2610.21765} {10.48550/arXiv.2610.21765} (\bibinfo {year} {2026})\BibitemShut {NoStop}%
\bibitem [{\citenamefont {Mayers}\ and\ \citenamefont {Yao}(2004)}]{MY04}%
  \BibitemOpen
  \bibfield  {author} {\bibinfo {author} {\bibfnamefont {D.}~\bibnamefont {Mayers}}\ and\ \bibinfo {author} {\bibfnamefont {A.}~\bibnamefont {Yao}},\ }\bibfield  {title} {\bibinfo {title} {{Self testing quantum apparatus}},\ }\href {https://doi.org/10.26421/QIC4.4-3} {\bibfield  {journal} {\bibinfo  {journal} {Quant. Inf. Comput.}\ }\textbf {\bibinfo {volume} {4}},\ \bibinfo {pages} {273} (\bibinfo {year} {2004})}\BibitemShut {NoStop}%
\bibitem [{\citenamefont {{\v{S}}upi{\'{c}}}\ and\ \citenamefont {Bowles}(2020)}]{SB20}%
  \BibitemOpen
  \bibfield  {author} {\bibinfo {author} {\bibfnamefont {I.}~\bibnamefont {{\v{S}}upi{\'{c}}}}\ and\ \bibinfo {author} {\bibfnamefont {J.}~\bibnamefont {Bowles}},\ }\bibfield  {title} {\bibinfo {title} {Self-testing of quantum systems: a review},\ }\href {https://doi.org/10.22331/q-2020-09-30-337} {\bibfield  {journal} {\bibinfo  {journal} {{Quantum}}\ }\textbf {\bibinfo {volume} {4}},\ \bibinfo {pages} {337} (\bibinfo {year} {2020})}\BibitemShut {NoStop}%
\bibitem [{\citenamefont {Kaniewski}(2017)}]{K17}%
  \BibitemOpen
  \bibfield  {author} {\bibinfo {author} {\bibfnamefont {J.}~\bibnamefont {Kaniewski}},\ }\bibfield  {title} {\bibinfo {title} {Self-testing of binary observables based on commutation},\ }\href {https://doi.org/10.1103/PhysRevA.95.062323} {\bibfield  {journal} {\bibinfo  {journal} {Phys. Rev. A}\ }\textbf {\bibinfo {volume} {95}},\ \bibinfo {pages} {062323} (\bibinfo {year} {2017})}\BibitemShut {NoStop}%
\bibitem [{\citenamefont {Sarkar}\ \emph {et~al.}(2026)\citenamefont {Sarkar}, \citenamefont {Orthey~Jr},\ and\ \citenamefont {Augusiak}}]{SOA26}%
  \BibitemOpen
  \bibfield  {author} {\bibinfo {author} {\bibfnamefont {S.}~\bibnamefont {Sarkar}}, \bibinfo {author} {\bibfnamefont {A.~C.}\ \bibnamefont {Orthey~Jr}},\ and\ \bibinfo {author} {\bibfnamefont {R.}~\bibnamefont {Augusiak}},\ }\bibfield  {title} {\bibinfo {title} {A universal scheme to self-test any quantum state or measurement},\ }\href {https://doi.org/10.1038/s41567-026-03181-y} {\bibfield  {journal} {\bibinfo  {journal} {Nat. Phys.}\ ,\ \bibinfo {pages} {1}} (\bibinfo {year} {2026})}\BibitemShut {NoStop}%
\bibitem [{\citenamefont {Agresti}\ \emph {et~al.}(2021)\citenamefont {Agresti}, \citenamefont {Polacchi}, \citenamefont {Poderini}, \citenamefont {Polino}, \citenamefont {Suprano}, \citenamefont {\ifmmode \check{S}\else \v{S}\fi{}upi\ifmmode~\acute{c}\else \'{c}\fi{}}, \citenamefont {Bowles}, \citenamefont {Carvacho}, \citenamefont {Cavalcanti},\ and\ \citenamefont {Sciarrino}}]{APPC21}%
  \BibitemOpen
  \bibfield  {author} {\bibinfo {author} {\bibfnamefont {I.}~\bibnamefont {Agresti}}, \bibinfo {author} {\bibfnamefont {B.}~\bibnamefont {Polacchi}}, \bibinfo {author} {\bibfnamefont {D.}~\bibnamefont {Poderini}}, \bibinfo {author} {\bibfnamefont {E.}~\bibnamefont {Polino}}, \bibinfo {author} {\bibfnamefont {A.}~\bibnamefont {Suprano}}, \bibinfo {author} {\bibfnamefont {I.}~\bibnamefont {\ifmmode \check{S}\else \v{S}\fi{}upi\ifmmode~\acute{c}\else \'{c}\fi{}}}, \bibinfo {author} {\bibfnamefont {J.}~\bibnamefont {Bowles}}, \bibinfo {author} {\bibfnamefont {G.}~\bibnamefont {Carvacho}}, \bibinfo {author} {\bibfnamefont {D.}~\bibnamefont {Cavalcanti}},\ and\ \bibinfo {author} {\bibfnamefont {F.}~\bibnamefont {Sciarrino}},\ }\bibfield  {title} {\bibinfo {title} {Experimental robust self-testing of the state generated by a quantum network},\ }\href {https://doi.org/10.1103/PRXQuantum.2.020346} {\bibfield  {journal} {\bibinfo  {journal} {PRX Quantum}\ }\textbf {\bibinfo {volume} {2}},\ \bibinfo {pages} {020346}
  (\bibinfo {year} {2021})}\BibitemShut {NoStop}%
\bibitem [{\citenamefont {Roy}\ and\ \citenamefont {Pan}(2023)}]{RP23}%
  \BibitemOpen
  \bibfield  {author} {\bibinfo {author} {\bibfnamefont {P.}~\bibnamefont {Roy}}\ and\ \bibinfo {author} {\bibfnamefont {A.~K.}\ \bibnamefont {Pan}},\ }\bibfield  {title} {\bibinfo {title} {Device-independent self-testing of unsharp measurements},\ }\href {https://doi.org/10.1088/1367-2630/acb4b5} {\bibfield  {journal} {\bibinfo  {journal} {New J. Phys.}\ }\textbf {\bibinfo {volume} {25}},\ \bibinfo {pages} {013040} (\bibinfo {year} {2023})}\BibitemShut {NoStop}%
\bibitem [{\citenamefont {Coladangelo}\ \emph {et~al.}(2017)\citenamefont {Coladangelo}, \citenamefont {Goh},\ and\ \citenamefont {Scarani}}]{CGS17}%
  \BibitemOpen
  \bibfield  {author} {\bibinfo {author} {\bibfnamefont {A.}~\bibnamefont {Coladangelo}}, \bibinfo {author} {\bibfnamefont {K.~T.}\ \bibnamefont {Goh}},\ and\ \bibinfo {author} {\bibfnamefont {V.}~\bibnamefont {Scarani}},\ }\bibfield  {title} {\bibinfo {title} {All pure bipartite entangled states can be self-tested},\ }\href {https://doi.org/10.1038/ncomms15485} {\bibfield  {journal} {\bibinfo  {journal} {Nat. Commun.}\ }\textbf {\bibinfo {volume} {8}},\ \bibinfo {pages} {15485} (\bibinfo {year} {2017})}\BibitemShut {NoStop}%
\bibitem [{\citenamefont {{B}ell}(1964)}]{Bell64}%
  \BibitemOpen
  \bibfield  {author} {\bibinfo {author} {\bibfnamefont {J.~S.}\ \bibnamefont {{B}ell}},\ }\bibfield  {title} {\bibinfo {title} {On the {E}instein {P}odolsky {R}osen paradox},\ }\href {https://doi.org/10.1103/PhysicsPhysiqueFizika.1.195} {\bibfield  {journal} {\bibinfo  {journal} {Phys. Phys. Fiz.}\ }\textbf {\bibinfo {volume} {1}},\ \bibinfo {pages} {195} (\bibinfo {year} {1964})}\BibitemShut {NoStop}%
\bibitem [{\citenamefont {Mckague}\ \emph {et~al.}(2012)\citenamefont {Mckague}, \citenamefont {Yang},\ and\ \citenamefont {Scarani}}]{M12}%
  \BibitemOpen
  \bibfield  {author} {\bibinfo {author} {\bibfnamefont {M.}~\bibnamefont {Mckague}}, \bibinfo {author} {\bibfnamefont {T.~H.}\ \bibnamefont {Yang}},\ and\ \bibinfo {author} {\bibfnamefont {V.}~\bibnamefont {Scarani}},\ }\bibfield  {title} {\bibinfo {title} {Robust self-testing of the singlet},\ }\href {https://api.semanticscholar.org/CorpusID:118535156} {\bibfield  {journal} {\bibinfo  {journal} {J. Phys. A: Math. Theor.}\ }\textbf {\bibinfo {volume} {45}} (\bibinfo {year} {2012})}\BibitemShut {NoStop}%
\bibitem [{\citenamefont {Šupić}\ \emph {et~al.}(2016)\citenamefont {Šupić}, \citenamefont {Augusiak}, \citenamefont {Salavrakos},\ and\ \citenamefont {Acín}}]{SASA16}%
  \BibitemOpen
  \bibfield  {author} {\bibinfo {author} {\bibfnamefont {I.}~\bibnamefont {Šupić}}, \bibinfo {author} {\bibfnamefont {R.}~\bibnamefont {Augusiak}}, \bibinfo {author} {\bibfnamefont {A.}~\bibnamefont {Salavrakos}},\ and\ \bibinfo {author} {\bibfnamefont {A.}~\bibnamefont {Acín}},\ }\bibfield  {title} {\bibinfo {title} {Self-testing protocols based on the chained {B}ell inequalities},\ }\href {https://doi.org/10.1088/1367-2630/18/3/035013} {\bibfield  {journal} {\bibinfo  {journal} {New J. Phys.}\ }\textbf {\bibinfo {volume} {18}},\ \bibinfo {pages} {035013} (\bibinfo {year} {2016})}\BibitemShut {NoStop}%
\bibitem [{\citenamefont {Zhang}\ \emph {et~al.}(2019{\natexlab{a}})\citenamefont {Zhang}, \citenamefont {Chen}, \citenamefont {Peng}, \citenamefont {Ye}, \citenamefont {Yin}, \citenamefont {Xu}, \citenamefont {Xu}, \citenamefont {Li},\ and\ \citenamefont {Guo}}]{ZCPG19}%
  \BibitemOpen
  \bibfield  {author} {\bibinfo {author} {\bibfnamefont {W.-H.}\ \bibnamefont {Zhang}}, \bibinfo {author} {\bibfnamefont {G.}~\bibnamefont {Chen}}, \bibinfo {author} {\bibfnamefont {X.-X.}\ \bibnamefont {Peng}}, \bibinfo {author} {\bibfnamefont {X.-J.}\ \bibnamefont {Ye}}, \bibinfo {author} {\bibfnamefont {P.}~\bibnamefont {Yin}}, \bibinfo {author} {\bibfnamefont {X.-Y.}\ \bibnamefont {Xu}}, \bibinfo {author} {\bibfnamefont {J.-S.}\ \bibnamefont {Xu}}, \bibinfo {author} {\bibfnamefont {C.-F.}\ \bibnamefont {Li}},\ and\ \bibinfo {author} {\bibfnamefont {G.-C.}\ \bibnamefont {Guo}},\ }\bibfield  {title} {\bibinfo {title} {Experimental realization of robust self-testing of {B}ell state measurements},\ }\href {https://doi.org/10.1103/PhysRevLett.122.090402} {\bibfield  {journal} {\bibinfo  {journal} {Phys. Rev. Lett.}\ }\textbf {\bibinfo {volume} {122}},\ \bibinfo {pages} {090402} (\bibinfo {year} {2019}{\natexlab{a}})}\BibitemShut {NoStop}%
\bibitem [{\citenamefont {Li}\ \emph {et~al.}(2022)\citenamefont {Li}, \citenamefont {Wang}, \citenamefont {Han},\ and\ \citenamefont {Zhu}}]{LWHZ22}%
  \BibitemOpen
  \bibfield  {author} {\bibinfo {author} {\bibfnamefont {X.}~\bibnamefont {Li}}, \bibinfo {author} {\bibfnamefont {Y.}~\bibnamefont {Wang}}, \bibinfo {author} {\bibfnamefont {Y.}~\bibnamefont {Han}},\ and\ \bibinfo {author} {\bibfnamefont {S.-N.}\ \bibnamefont {Zhu}},\ }\bibfield  {title} {\bibinfo {title} {Self-testing of different entanglement resources via fixed measurement settings},\ }\href {https://doi.org/10.1103/PhysRevA.106.052418} {\bibfield  {journal} {\bibinfo  {journal} {Phys. Rev. A}\ }\textbf {\bibinfo {volume} {106}},\ \bibinfo {pages} {052418} (\bibinfo {year} {2022})}\BibitemShut {NoStop}%
\bibitem [{\citenamefont {Singh}\ and\ \citenamefont {Cabello}(2024)}]{JC24}%
  \BibitemOpen
  \bibfield  {author} {\bibinfo {author} {\bibfnamefont {J.}~\bibnamefont {Singh}}\ and\ \bibinfo {author} {\bibfnamefont {A.}~\bibnamefont {Cabello}},\ }\bibfield  {title} {\bibinfo {title} {Loophole-free {B}ell tests with randomly chosen subsets of measurement settings},\ }\href {https://doi.org/10.1103/PhysRevA.109.022204} {\bibfield  {journal} {\bibinfo  {journal} {Phys. Rev. A}\ }\textbf {\bibinfo {volume} {109}},\ \bibinfo {pages} {022204} (\bibinfo {year} {2024})}\BibitemShut {NoStop}%
\bibitem [{\citenamefont {Mehta}\ \emph {et~al.}(2025)\citenamefont {Mehta}, \citenamefont {Paddock},\ and\ \citenamefont {Wooltorton}}]{MPW25}%
  \BibitemOpen
  \bibfield  {author} {\bibinfo {author} {\bibfnamefont {A.}~\bibnamefont {Mehta}}, \bibinfo {author} {\bibfnamefont {C.}~\bibnamefont {Paddock}},\ and\ \bibinfo {author} {\bibfnamefont {L.}~\bibnamefont {Wooltorton}},\ }\bibfield  {title} {\bibinfo {title} {{Self-testing in the compiled setting via tilted-{CHSH} inequalities}},\ }in\ \href {https://doi.org/10.4230/LIPIcs.TQC.2025.8} {\emph {\bibinfo {booktitle} {20th Conference on the Theory of Quantum Computation, Communication and Cryptography (TQC 2025)}}},\ \bibinfo {series} {Leibniz International Proceedings in Informatics (LIPIcs)}, Vol.\ \bibinfo {volume} {350},\ \bibinfo {editor} {edited by\ \bibinfo {editor} {\bibfnamefont {B.}~\bibnamefont {Fefferman}}}\ (\bibinfo  {publisher} {Schloss Dagstuhl -- Leibniz-Zentrum f{\"u}r Informatik},\ \bibinfo {address} {Dagstuhl, Germany},\ \bibinfo {year} {2025})\ pp.\ \bibinfo {pages} {8:1--8:19}\BibitemShut {NoStop}%
\bibitem [{\citenamefont {Kaniewski}(2016)}]{K16}%
  \BibitemOpen
  \bibfield  {author} {\bibinfo {author} {\bibfnamefont {J.}~\bibnamefont {Kaniewski}},\ }\bibfield  {title} {\bibinfo {title} {Analytic and nearly optimal self-testing bounds for the {C}lauser-{H}orne-{S}himony-{H}olt and {M}ermin inequalities},\ }\href {https://doi.org/10.1103/PhysRevLett.117.070402} {\bibfield  {journal} {\bibinfo  {journal} {Phys. Rev. Lett.}\ }\textbf {\bibinfo {volume} {117}},\ \bibinfo {pages} {070402} (\bibinfo {year} {2016})}\BibitemShut {NoStop}%
\bibitem [{\citenamefont {Bao}\ \emph {et~al.}(2025)\citenamefont {Bao}, \citenamefont {Zhao}, \citenamefont {Zhou}, \citenamefont {Xu}, \citenamefont {Huang},\ and\ \citenamefont {Tan}}]{BZZXHT25}%
  \BibitemOpen
  \bibfield  {author} {\bibinfo {author} {\bibfnamefont {D.}~\bibnamefont {Bao}}, \bibinfo {author} {\bibfnamefont {Z.}~\bibnamefont {Zhao}}, \bibinfo {author} {\bibfnamefont {Y.}~\bibnamefont {Zhou}}, \bibinfo {author} {\bibfnamefont {Q.}~\bibnamefont {Xu}}, \bibinfo {author} {\bibfnamefont {R.}~\bibnamefont {Huang}},\ and\ \bibinfo {author} {\bibfnamefont {X.}~\bibnamefont {Tan}},\ }\bibfield  {title} {\bibinfo {title} {Graph-theoretic approach for self-testing of multipartite {M}ermin-{A}rdehali-{B}elinski\ifmmode \check{i}\else \v{i}\fi{}-{K}lyshko inequalities},\ }\href {https://doi.org/10.1103/w3zz-xps6} {\bibfield  {journal} {\bibinfo  {journal} {Phys. Rev. A}\ }\textbf {\bibinfo {volume} {112}},\ \bibinfo {pages} {052455} (\bibinfo {year} {2025})}\BibitemShut {NoStop}%
\bibitem [{\citenamefont {Rai}\ \emph {et~al.}(2022)\citenamefont {Rai}, \citenamefont {Pivoluska}, \citenamefont {Sasmal}, \citenamefont {Banik}, \citenamefont {Ghosh},\ and\ \citenamefont {Plesch}}]{AMSMSM22}%
  \BibitemOpen
  \bibfield  {author} {\bibinfo {author} {\bibfnamefont {A.}~\bibnamefont {Rai}}, \bibinfo {author} {\bibfnamefont {M.}~\bibnamefont {Pivoluska}}, \bibinfo {author} {\bibfnamefont {S.}~\bibnamefont {Sasmal}}, \bibinfo {author} {\bibfnamefont {M.}~\bibnamefont {Banik}}, \bibinfo {author} {\bibfnamefont {S.}~\bibnamefont {Ghosh}},\ and\ \bibinfo {author} {\bibfnamefont {M.}~\bibnamefont {Plesch}},\ }\bibfield  {title} {\bibinfo {title} {Self-testing quantum states via nonmaximal violation in {H}ardy's test of nonlocality},\ }\href {https://doi.org/10.1103/PhysRevA.105.052227} {\bibfield  {journal} {\bibinfo  {journal} {Phys. Rev. A}\ }\textbf {\bibinfo {volume} {105}},\ \bibinfo {pages} {052227} (\bibinfo {year} {2022})}\BibitemShut {NoStop}%
\bibitem [{\citenamefont {Spekkens}(2005)}]{S05}%
  \BibitemOpen
  \bibfield  {author} {\bibinfo {author} {\bibfnamefont {R.~W.}\ \bibnamefont {Spekkens}},\ }\bibfield  {title} {\bibinfo {title} {Contextuality for preparations, transformations, and unsharp measurements},\ }\href {https://doi.org/10.1103/PhysRevA.71.052108} {\bibfield  {journal} {\bibinfo  {journal} {Phys. Rev. A}\ }\textbf {\bibinfo {volume} {71}},\ \bibinfo {pages} {052108} (\bibinfo {year} {2005})}\BibitemShut {NoStop}%
\bibitem [{\citenamefont {Hazra}\ \emph {et~al.}(2026)\citenamefont {Hazra}, \citenamefont {Saha}, \citenamefont {Chaturvedi}, \citenamefont {Bera},\ and\ \citenamefont {Majumdar}}]{SDASA26}%
  \BibitemOpen
  \bibfield  {author} {\bibinfo {author} {\bibfnamefont {S.}~\bibnamefont {Hazra}}, \bibinfo {author} {\bibfnamefont {D.}~\bibnamefont {Saha}}, \bibinfo {author} {\bibfnamefont {A.}~\bibnamefont {Chaturvedi}}, \bibinfo {author} {\bibfnamefont {S.}~\bibnamefont {Bera}},\ and\ \bibinfo {author} {\bibfnamefont {A.~S.}\ \bibnamefont {Majumdar}},\ }\bibfield  {title} {\bibinfo {title} {Efficient {C}omputation of {G}eneralized {N}oncontextual {P}olytopes and {Q}uantum violation of their {F}acet {I}nequalities},\ }\href {https://doi.org/10.22331/q-2026-03-09-2015} {\bibfield  {journal} {\bibinfo  {journal} {{Quantum}}\ }\textbf {\bibinfo {volume} {10}},\ \bibinfo {pages} {2015} (\bibinfo {year} {2026})}\BibitemShut {NoStop}%
\bibitem [{\citenamefont {Leggett}\ and\ \citenamefont {Garg}(1985)}]{LG85}%
  \BibitemOpen
  \bibfield  {author} {\bibinfo {author} {\bibfnamefont {A.~J.}\ \bibnamefont {Leggett}}\ and\ \bibinfo {author} {\bibfnamefont {A.}~\bibnamefont {Garg}},\ }\bibfield  {title} {\bibinfo {title} {Quantum mechanics versus macroscopic realism: Is the flux there when nobody looks?},\ }\href {https://doi.org/10.1103/PhysRevLett.54.857} {\bibfield  {journal} {\bibinfo  {journal} {Phys. Rev. Lett.}\ }\textbf {\bibinfo {volume} {54}},\ \bibinfo {pages} {857} (\bibinfo {year} {1985})}\BibitemShut {NoStop}%
\bibitem [{\citenamefont {Bharti}\ \emph {et~al.}(2019{\natexlab{a}})\citenamefont {Bharti}, \citenamefont {Ray}, \citenamefont {Varvitsiotis}, \citenamefont {Warsi}, \citenamefont {Cabello},\ and\ \citenamefont {Kwek}}]{KBMA19}%
  \BibitemOpen
  \bibfield  {author} {\bibinfo {author} {\bibfnamefont {K.}~\bibnamefont {Bharti}}, \bibinfo {author} {\bibfnamefont {M.}~\bibnamefont {Ray}}, \bibinfo {author} {\bibfnamefont {A.}~\bibnamefont {Varvitsiotis}}, \bibinfo {author} {\bibfnamefont {N.~A.}\ \bibnamefont {Warsi}}, \bibinfo {author} {\bibfnamefont {A.}~\bibnamefont {Cabello}},\ and\ \bibinfo {author} {\bibfnamefont {L.-C.}\ \bibnamefont {Kwek}},\ }\bibfield  {title} {\bibinfo {title} {Robust self-testing of quantum systems via noncontextuality inequalities},\ }\href {https://doi.org/10.1103/PhysRevLett.122.250403} {\bibfield  {journal} {\bibinfo  {journal} {Phys. Rev. Lett.}\ }\textbf {\bibinfo {volume} {122}},\ \bibinfo {pages} {250403} (\bibinfo {year} {2019}{\natexlab{a}})}\BibitemShut {NoStop}%
\bibitem [{\citenamefont {Paul}\ \emph {et~al.}(2024)\citenamefont {Paul}, \citenamefont {Sasmal},\ and\ \citenamefont {Pan}}]{PSP24}%
  \BibitemOpen
  \bibfield  {author} {\bibinfo {author} {\bibfnamefont {R.}~\bibnamefont {Paul}}, \bibinfo {author} {\bibfnamefont {S.}~\bibnamefont {Sasmal}},\ and\ \bibinfo {author} {\bibfnamefont {A.~K.}\ \bibnamefont {Pan}},\ }\bibfield  {title} {\bibinfo {title} {Self-testing of multiple unsharpness parameters through sequential violations of a noncontextual inequality},\ }\href {https://doi.org/10.1103/PhysRevA.110.012444} {\bibfield  {journal} {\bibinfo  {journal} {Phys. Rev. A}\ }\textbf {\bibinfo {volume} {110}},\ \bibinfo {pages} {012444} (\bibinfo {year} {2024})}\BibitemShut {NoStop}%
\bibitem [{\citenamefont {Singh}\ \emph {et~al.}(2024)\citenamefont {Singh}, \citenamefont {Foreman}, \citenamefont {Bharti},\ and\ \citenamefont {Cabello}}]{JFBC24}%
  \BibitemOpen
  \bibfield  {author} {\bibinfo {author} {\bibfnamefont {J.}~\bibnamefont {Singh}}, \bibinfo {author} {\bibfnamefont {C.}~\bibnamefont {Foreman}}, \bibinfo {author} {\bibfnamefont {K.}~\bibnamefont {Bharti}},\ and\ \bibinfo {author} {\bibfnamefont {A.}~\bibnamefont {Cabello}},\ }\bibfield  {title} {\bibinfo {title} {Local contextuality-based self-tests are sufficient for randomness expansion secure against quantum adversaries},\ }\bibfield  {journal} {\bibinfo  {journal} {arXiv}\ }\href {https://doi.org/10.48550/arXiv.2409.20082} {10.48550/arXiv.2409.20082} (\bibinfo {year} {2024}),\ \Eprint {https://arxiv.org/abs/2409.20082} {2409.20082} \BibitemShut {NoStop}%
\bibitem [{\citenamefont {Saha}\ \emph {et~al.}(2020)\citenamefont {Saha}, \citenamefont {Santos},\ and\ \citenamefont {Augusiak}}]{SSA20}%
  \BibitemOpen
  \bibfield  {author} {\bibinfo {author} {\bibfnamefont {D.}~\bibnamefont {Saha}}, \bibinfo {author} {\bibfnamefont {R.}~\bibnamefont {Santos}},\ and\ \bibinfo {author} {\bibfnamefont {R.}~\bibnamefont {Augusiak}},\ }\bibfield  {title} {\bibinfo {title} {Sum-of-squares decompositions for a family of noncontextuality inequalities and self-testing of quantum devices},\ }\href {https://doi.org/10.22331/q-2020-08-03-302} {\bibfield  {journal} {\bibinfo  {journal} {{Quantum}}\ }\textbf {\bibinfo {volume} {4}},\ \bibinfo {pages} {302} (\bibinfo {year} {2020})}\BibitemShut {NoStop}%
\bibitem [{\citenamefont {Hu}\ \emph {et~al.}(2023)\citenamefont {Hu}, \citenamefont {Xie}, \citenamefont {Arora}, \citenamefont {Ai}, \citenamefont {Bharti}, \citenamefont {Zhang}, \citenamefont {Wu}, \citenamefont {Chen}, \citenamefont {Cui}, \citenamefont {Liu}, \citenamefont {Huang}, \citenamefont {Li}, \citenamefont {Guo}, \citenamefont {Roland}, \citenamefont {Cabello},\ and\ \citenamefont {Kwek}}]{HK23}%
  \BibitemOpen
  \bibfield  {author} {\bibinfo {author} {\bibfnamefont {X.-M.}\ \bibnamefont {Hu}}, \bibinfo {author} {\bibfnamefont {Y.}~\bibnamefont {Xie}}, \bibinfo {author} {\bibfnamefont {A.~S.}\ \bibnamefont {Arora}}, \bibinfo {author} {\bibfnamefont {M.-Z.}\ \bibnamefont {Ai}}, \bibinfo {author} {\bibfnamefont {K.}~\bibnamefont {Bharti}}, \bibinfo {author} {\bibfnamefont {J.}~\bibnamefont {Zhang}}, \bibinfo {author} {\bibfnamefont {W.}~\bibnamefont {Wu}}, \bibinfo {author} {\bibfnamefont {P.-X.}\ \bibnamefont {Chen}}, \bibinfo {author} {\bibfnamefont {J.-M.}\ \bibnamefont {Cui}}, \bibinfo {author} {\bibfnamefont {B.-H.}\ \bibnamefont {Liu}}, \bibinfo {author} {\bibfnamefont {Y.-F.}\ \bibnamefont {Huang}}, \bibinfo {author} {\bibfnamefont {C.-F.}\ \bibnamefont {Li}}, \bibinfo {author} {\bibfnamefont {G.-C.}\ \bibnamefont {Guo}}, \bibinfo {author} {\bibfnamefont {J.}~\bibnamefont {Roland}}, \bibinfo {author} {\bibfnamefont {A.}~\bibnamefont {Cabello}},\ and\ \bibinfo {author} {\bibfnamefont {L.-C.}\ \bibnamefont
  {Kwek}},\ }\bibfield  {title} {\bibinfo {title} {Self-testing of a single quantum system from theory to experiment},\ }\href {https://doi.org/10.1038/s41534-023-00769-7} {\bibfield  {journal} {\bibinfo  {journal} {Npj Quantum Inf.}\ }\textbf {\bibinfo {volume} {9}},\ \bibinfo {pages} {103} (\bibinfo {year} {2023})}\BibitemShut {NoStop}%
\bibitem [{\citenamefont {Maity}\ \emph {et~al.}(2021)\citenamefont {Maity}, \citenamefont {Mal}, \citenamefont {Jebarathinam},\ and\ \citenamefont {Majumdar}}]{ASCA21}%
  \BibitemOpen
  \bibfield  {author} {\bibinfo {author} {\bibfnamefont {A.~G.}\ \bibnamefont {Maity}}, \bibinfo {author} {\bibfnamefont {S.}~\bibnamefont {Mal}}, \bibinfo {author} {\bibfnamefont {C.}~\bibnamefont {Jebarathinam}},\ and\ \bibinfo {author} {\bibfnamefont {A.~S.}\ \bibnamefont {Majumdar}},\ }\bibfield  {title} {\bibinfo {title} {Self-testing of binary {P}auli measurements requiring neither entanglement nor any dimensional restriction},\ }\href {https://doi.org/10.1103/PhysRevA.103.062604} {\bibfield  {journal} {\bibinfo  {journal} {Phys. Rev. A}\ }\textbf {\bibinfo {volume} {103}},\ \bibinfo {pages} {062604} (\bibinfo {year} {2021})}\BibitemShut {NoStop}%
\bibitem [{\citenamefont {Bharti}\ \emph {et~al.}(2022)\citenamefont {Bharti}, \citenamefont {Ray}, \citenamefont {Xu}, \citenamefont {Hayashi}, \citenamefont {Kwek},\ and\ \citenamefont {Cabello}}]{KBMZ22}%
  \BibitemOpen
  \bibfield  {author} {\bibinfo {author} {\bibfnamefont {K.}~\bibnamefont {Bharti}}, \bibinfo {author} {\bibfnamefont {M.}~\bibnamefont {Ray}}, \bibinfo {author} {\bibfnamefont {Z.-P.}\ \bibnamefont {Xu}}, \bibinfo {author} {\bibfnamefont {M.}~\bibnamefont {Hayashi}}, \bibinfo {author} {\bibfnamefont {L.-C.}\ \bibnamefont {Kwek}},\ and\ \bibinfo {author} {\bibfnamefont {A.}~\bibnamefont {Cabello}},\ }\bibfield  {title} {\bibinfo {title} {Graph-theoretic approach for self-testing in {B}ell scenarios},\ }\href {https://doi.org/10.1103/PRXQuantum.3.030344} {\bibfield  {journal} {\bibinfo  {journal} {PRX Quantum}\ }\textbf {\bibinfo {volume} {3}},\ \bibinfo {pages} {030344} (\bibinfo {year} {2022})}\BibitemShut {NoStop}%
\bibitem [{\citenamefont {Brunner}\ \emph {et~al.}(2014)\citenamefont {Brunner}, \citenamefont {Cavalcanti}, \citenamefont {Pironio}, \citenamefont {Scarani},\ and\ \citenamefont {Wehner}}]{BCPSW14}%
  \BibitemOpen
  \bibfield  {author} {\bibinfo {author} {\bibfnamefont {N.}~\bibnamefont {Brunner}}, \bibinfo {author} {\bibfnamefont {D.}~\bibnamefont {Cavalcanti}}, \bibinfo {author} {\bibfnamefont {S.}~\bibnamefont {Pironio}}, \bibinfo {author} {\bibfnamefont {V.}~\bibnamefont {Scarani}},\ and\ \bibinfo {author} {\bibfnamefont {S.}~\bibnamefont {Wehner}},\ }\bibfield  {title} {\bibinfo {title} {{B}ell nonlocality},\ }\href {https://doi.org/10.1103/RevModPhys.86.419} {\bibfield  {journal} {\bibinfo  {journal} {Rev. Mod. Phys.}\ }\textbf {\bibinfo {volume} {86}},\ \bibinfo {pages} {419} (\bibinfo {year} {2014})}\BibitemShut {NoStop}%
\bibitem [{\citenamefont {Bowles}\ \emph {et~al.}(2018)\citenamefont {Bowles}, \citenamefont {\ifmmode \check{S}\else \v{S}\fi{}upi\ifmmode~\acute{c}\else \'{c}\fi{}}, \citenamefont {Cavalcanti},\ and\ \citenamefont {Ac\'{\i}n}}]{BSCA18}%
  \BibitemOpen
  \bibfield  {author} {\bibinfo {author} {\bibfnamefont {J.}~\bibnamefont {Bowles}}, \bibinfo {author} {\bibfnamefont {I.}~\bibnamefont {\ifmmode \check{S}\else \v{S}\fi{}upi\ifmmode~\acute{c}\else \'{c}\fi{}}}, \bibinfo {author} {\bibfnamefont {D.}~\bibnamefont {Cavalcanti}},\ and\ \bibinfo {author} {\bibfnamefont {A.}~\bibnamefont {Ac\'{\i}n}},\ }\bibfield  {title} {\bibinfo {title} {Self-testing of {P}auli observables for device-independent entanglement certification},\ }\href {https://doi.org/10.1103/PhysRevA.98.042336} {\bibfield  {journal} {\bibinfo  {journal} {Phys. Rev. A}\ }\textbf {\bibinfo {volume} {98}},\ \bibinfo {pages} {042336} (\bibinfo {year} {2018})}\BibitemShut {NoStop}%
\bibitem [{\citenamefont {Wu}\ \emph {et~al.}(2016)\citenamefont {Wu}, \citenamefont {Bancal}, \citenamefont {McKague},\ and\ \citenamefont {Scarani}}]{WBDMS16}%
  \BibitemOpen
  \bibfield  {author} {\bibinfo {author} {\bibfnamefont {X.}~\bibnamefont {Wu}}, \bibinfo {author} {\bibfnamefont {J.-D.}\ \bibnamefont {Bancal}}, \bibinfo {author} {\bibfnamefont {M.}~\bibnamefont {McKague}},\ and\ \bibinfo {author} {\bibfnamefont {V.}~\bibnamefont {Scarani}},\ }\bibfield  {title} {\bibinfo {title} {Device-independent parallel self-testing of two singlets},\ }\href {https://doi.org/10.1103/PhysRevA.93.062121} {\bibfield  {journal} {\bibinfo  {journal} {Phys. Rev. A}\ }\textbf {\bibinfo {volume} {93}},\ \bibinfo {pages} {062121} (\bibinfo {year} {2016})}\BibitemShut {NoStop}%
\bibitem [{\citenamefont {Ac{\'\i}n}\ \emph {et~al.}(2018)\citenamefont {Ac{\'\i}n}, \citenamefont {Bloch}, \citenamefont {Buhrman}, \citenamefont {Calarco}, \citenamefont {Eichler}, \citenamefont {Eisert}, \citenamefont {Esteve}, \citenamefont {Gisin}, \citenamefont {Glaser}, \citenamefont {Jelezko}, \citenamefont {Kuhr}, \citenamefont {Lewenstein}, \citenamefont {Riedel}, \citenamefont {Schmidt}, \citenamefont {Thew}, \citenamefont {Wallraff}, \citenamefont {Walmsley},\ and\ \citenamefont {Wilhelm}}]{ABB18}%
  \BibitemOpen
  \bibfield  {author} {\bibinfo {author} {\bibfnamefont {A.}~\bibnamefont {Ac{\'\i}n}}, \bibinfo {author} {\bibfnamefont {I.}~\bibnamefont {Bloch}}, \bibinfo {author} {\bibfnamefont {H.}~\bibnamefont {Buhrman}}, \bibinfo {author} {\bibfnamefont {T.}~\bibnamefont {Calarco}}, \bibinfo {author} {\bibfnamefont {C.}~\bibnamefont {Eichler}}, \bibinfo {author} {\bibfnamefont {J.}~\bibnamefont {Eisert}}, \bibinfo {author} {\bibfnamefont {D.}~\bibnamefont {Esteve}}, \bibinfo {author} {\bibfnamefont {N.}~\bibnamefont {Gisin}}, \bibinfo {author} {\bibfnamefont {S.~J.}\ \bibnamefont {Glaser}}, \bibinfo {author} {\bibfnamefont {F.}~\bibnamefont {Jelezko}}, \bibinfo {author} {\bibfnamefont {S.}~\bibnamefont {Kuhr}}, \bibinfo {author} {\bibfnamefont {M.}~\bibnamefont {Lewenstein}}, \bibinfo {author} {\bibfnamefont {M.~F.}\ \bibnamefont {Riedel}}, \bibinfo {author} {\bibfnamefont {P.~O.}\ \bibnamefont {Schmidt}}, \bibinfo {author} {\bibfnamefont {R.}~\bibnamefont {Thew}}, \bibinfo {author} {\bibfnamefont {A.}~\bibnamefont
  {Wallraff}}, \bibinfo {author} {\bibfnamefont {I.}~\bibnamefont {Walmsley}},\ and\ \bibinfo {author} {\bibfnamefont {F.~K.}\ \bibnamefont {Wilhelm}},\ }\bibfield  {title} {\bibinfo {title} {The quantum technologies roadmap: a {E}uropean community view},\ }\href {https://doi.org/10.1088/1367-2630/aad1ea} {\bibfield  {journal} {\bibinfo  {journal} {New J. Phys.}\ }\textbf {\bibinfo {volume} {20}},\ \bibinfo {pages} {080201} (\bibinfo {year} {2018})}\BibitemShut {NoStop}%
\bibitem [{\citenamefont {Preskill}(2018)}]{P18}%
  \BibitemOpen
  \bibfield  {author} {\bibinfo {author} {\bibfnamefont {J.}~\bibnamefont {Preskill}},\ }\bibfield  {title} {\bibinfo {title} {Quantum {C}omputing in the {NISQ} era and beyond},\ }\href {https://doi.org/10.22331/q-2018-08-06-79} {\bibfield  {journal} {\bibinfo  {journal} {{Quantum}}\ }\textbf {\bibinfo {volume} {2}},\ \bibinfo {pages} {79} (\bibinfo {year} {2018})}\BibitemShut {NoStop}%
\bibitem [{\citenamefont {Banaszek}\ \emph {et~al.}(2013)\citenamefont {Banaszek}, \citenamefont {Cramer},\ and\ \citenamefont {Gross}}]{BCG13}%
  \BibitemOpen
  \bibfield  {author} {\bibinfo {author} {\bibfnamefont {K.}~\bibnamefont {Banaszek}}, \bibinfo {author} {\bibfnamefont {M.}~\bibnamefont {Cramer}},\ and\ \bibinfo {author} {\bibfnamefont {D.}~\bibnamefont {Gross}},\ }\bibfield  {title} {\bibinfo {title} {Focus on quantum tomography},\ }\href {https://doi.org/10.1088/1367-2630/15/12/125020} {\bibfield  {journal} {\bibinfo  {journal} {New J. Phys.}\ }\textbf {\bibinfo {volume} {15}},\ \bibinfo {pages} {125020} (\bibinfo {year} {2013})}\BibitemShut {NoStop}%
\bibitem [{\citenamefont {Palmieri}\ \emph {et~al.}(2024)\citenamefont {Palmieri}, \citenamefont {M\"uller-Rigat}, \citenamefont {Srivastava}, \citenamefont {Lewenstein}, \citenamefont {Rajchel-Mieldzio\ifmmode~\acute{c}\else \'{c}\fi{}},\ and\ \citenamefont {P\l{}odzie\ifmmode~\acute{n}\else \'{n}\fi{}}}]{PMSLP24}%
  \BibitemOpen
  \bibfield  {author} {\bibinfo {author} {\bibfnamefont {A.~M.}\ \bibnamefont {Palmieri}}, \bibinfo {author} {\bibfnamefont {G.}~\bibnamefont {M\"uller-Rigat}}, \bibinfo {author} {\bibfnamefont {A.~K.}\ \bibnamefont {Srivastava}}, \bibinfo {author} {\bibfnamefont {M.}~\bibnamefont {Lewenstein}}, \bibinfo {author} {\bibfnamefont {G.}~\bibnamefont {Rajchel-Mieldzio\ifmmode~\acute{c}\else \'{c}\fi{}}},\ and\ \bibinfo {author} {\bibfnamefont {M.}~\bibnamefont {P\l{}odzie\ifmmode~\acute{n}\else \'{n}\fi{}}},\ }\bibfield  {title} {\bibinfo {title} {Enhancing quantum state tomography via resource-efficient attention-based neural networks},\ }\href {https://doi.org/10.1103/PhysRevResearch.6.033248} {\bibfield  {journal} {\bibinfo  {journal} {Phys. Rev. Res.}\ }\textbf {\bibinfo {volume} {6}},\ \bibinfo {pages} {033248} (\bibinfo {year} {2024})}\BibitemShut {NoStop}%
\bibitem [{\citenamefont {Mauro~D’Ariano}\ \emph {et~al.}(2003)\citenamefont {Mauro~D’Ariano}, \citenamefont {Paris},\ and\ \citenamefont {Sacchi}}]{MPS03}%
  \BibitemOpen
  \bibfield  {author} {\bibinfo {author} {\bibfnamefont {G.}~\bibnamefont {Mauro~D’Ariano}}, \bibinfo {author} {\bibfnamefont {M.~G.}\ \bibnamefont {Paris}},\ and\ \bibinfo {author} {\bibfnamefont {M.~F.}\ \bibnamefont {Sacchi}},\ }\bibfield  {title} {\bibinfo {title} {Quantum tomography},\ }\bibfield  {booktitle} {\emph {\bibinfo {booktitle} {Advances in imaging and electron physics}},\ }\href {https://doi.org/10.1016/s1076-5670(03)80065-4} {\bibfield  {journal} {\bibinfo  {journal} {arXiv}\ ,\ \bibinfo {pages} {205–308}} (\bibinfo {year} {2003})}\BibitemShut {NoStop}%
\bibitem [{\citenamefont {Reichardt}\ \emph {et~al.}(2013)\citenamefont {Reichardt}, \citenamefont {Unger},\ and\ \citenamefont {Vazirani}}]{RFU13}%
  \BibitemOpen
  \bibfield  {author} {\bibinfo {author} {\bibfnamefont {B.~W.}\ \bibnamefont {Reichardt}}, \bibinfo {author} {\bibfnamefont {F.}~\bibnamefont {Unger}},\ and\ \bibinfo {author} {\bibfnamefont {U.}~\bibnamefont {Vazirani}},\ }\bibfield  {title} {\bibinfo {title} {Classical command of quantum systems},\ }\href {https://doi.org/10.1038/nature12035} {\bibfield  {journal} {\bibinfo  {journal} {Nature}\ }\textbf {\bibinfo {volume} {496}},\ \bibinfo {pages} {456} (\bibinfo {year} {2013})}\BibitemShut {NoStop}%
\bibitem [{\citenamefont {Coladangelo}(2017)}]{C17}%
  \BibitemOpen
  \bibfield  {author} {\bibinfo {author} {\bibfnamefont {A.~W.}\ \bibnamefont {Coladangelo}},\ }\bibfield  {title} {\bibinfo {title} {{Parallel self-testing of (tilted) EPR pairs via copies of (tilted) CHSH and the magic square game}},\ }\href {https://doi.org/10.26421/QIC17.9-10-6} {\bibfield  {journal} {\bibinfo  {journal} {Quant. Inf. Comput.}\ }\textbf {\bibinfo {volume} {17}},\ \bibinfo {pages} {0831} (\bibinfo {year} {2017})},\ \Eprint {https://arxiv.org/abs/1609.03687} {arXiv:1609.03687 [quant-ph]} \BibitemShut {NoStop}%
\bibitem [{\citenamefont {McKague}(2017)}]{M17}%
  \BibitemOpen
  \bibfield  {author} {\bibinfo {author} {\bibfnamefont {M.}~\bibnamefont {McKague}},\ }\bibfield  {title} {\bibinfo {title} {Self-testing in parallel with {CHSH}},\ }\href {https://doi.org/10.22331/q-2017-04-25-1} {\bibfield  {journal} {\bibinfo  {journal} {{Quantum}}\ }\textbf {\bibinfo {volume} {1}},\ \bibinfo {pages} {1} (\bibinfo {year} {2017})}\BibitemShut {NoStop}%
\bibitem [{\citenamefont {McKague}(2016)}]{M16}%
  \BibitemOpen
  \bibfield  {author} {\bibinfo {author} {\bibfnamefont {M.}~\bibnamefont {McKague}},\ }\bibfield  {title} {\bibinfo {title} {Self-testing in parallel},\ }\href {https://doi.org/10.1088/1367-2630/18/4/045013} {\bibfield  {journal} {\bibinfo  {journal} {New J. Phys.}\ }\textbf {\bibinfo {volume} {18}},\ \bibinfo {pages} {045013} (\bibinfo {year} {2016})}\BibitemShut {NoStop}%
\bibitem [{\citenamefont {Breiner}\ \emph {et~al.}(2019)\citenamefont {Breiner}, \citenamefont {Kalev},\ and\ \citenamefont {Miller}}]{BKM19}%
  \BibitemOpen
  \bibfield  {author} {\bibinfo {author} {\bibfnamefont {S.}~\bibnamefont {Breiner}}, \bibinfo {author} {\bibfnamefont {A.}~\bibnamefont {Kalev}},\ and\ \bibinfo {author} {\bibfnamefont {C.~A.}\ \bibnamefont {Miller}},\ }\bibfield  {title} {\bibinfo {title} {Parallel self-testing of the {G}{H}{Z} state with a proof by diagrams},\ }\href {https://doi.org/10.4204/eptcs.287.3} {\bibfield  {journal} {\bibinfo  {journal} {Electron. proc. theor. comput. sci.}\ }\textbf {\bibinfo {volume} {287}},\ \bibinfo {pages} {43–66} (\bibinfo {year} {2019})}\BibitemShut {NoStop}%
\bibitem [{\citenamefont {Coudron}\ and\ \citenamefont {Natarajan}(2016)}]{CN16}%
  \BibitemOpen
  \bibfield  {author} {\bibinfo {author} {\bibfnamefont {M.}~\bibnamefont {Coudron}}\ and\ \bibinfo {author} {\bibfnamefont {A.}~\bibnamefont {Natarajan}},\ }\bibfield  {title} {\bibinfo {title} {The parallel-repeated magic square game is rigid},\ }\bibfield  {journal} {\bibinfo  {journal} {arXiv}\ }\href {https://doi.org/10.48550/arXiv.1609.06306} {10.48550/arXiv.1609.06306} (\bibinfo {year} {2016}),\ \Eprint {https://arxiv.org/abs/1609.06306} {arXiv:1609.06306 [quant-ph]} \BibitemShut {NoStop}%
\bibitem [{\citenamefont {Adamson}\ and\ \citenamefont {Wallden}(2022)}]{AW22}%
  \BibitemOpen
  \bibfield  {author} {\bibinfo {author} {\bibfnamefont {S.~A.}\ \bibnamefont {Adamson}}\ and\ \bibinfo {author} {\bibfnamefont {P.}~\bibnamefont {Wallden}},\ }\bibfield  {title} {\bibinfo {title} {Practical parallel self-testing of {B}ell states via magic rectangles},\ }\href {https://doi.org/10.1103/PhysRevA.105.032456} {\bibfield  {journal} {\bibinfo  {journal} {Phys. Rev. A}\ }\textbf {\bibinfo {volume} {105}},\ \bibinfo {pages} {032456} (\bibinfo {year} {2022})}\BibitemShut {NoStop}%
\bibitem [{\citenamefont {Kalev}\ and\ \citenamefont {Miller}(2018)}]{KM18}%
  \BibitemOpen
  \bibfield  {author} {\bibinfo {author} {\bibfnamefont {A.}~\bibnamefont {Kalev}}\ and\ \bibinfo {author} {\bibfnamefont {C.~A.}\ \bibnamefont {Miller}},\ }\bibfield  {title} {\bibinfo {title} {Rigidity of the magic pentagram game},\ }\href {https://doi.org/10.1088/2058-9565/aa931d} {\bibfield  {journal} {\bibinfo  {journal} {Quantum Sci. Technol.}\ }\textbf {\bibinfo {volume} {3}},\ \bibinfo {pages} {015002} (\bibinfo {year} {2018})}\BibitemShut {NoStop}%
\bibitem [{\citenamefont {Coladangelo}\ \emph {et~al.}(2020)\citenamefont {Coladangelo}, \citenamefont {Grilo}, \citenamefont {Jeffery},\ and\ \citenamefont {Vidick}}]{CGJV20}%
  \BibitemOpen
  \bibfield  {author} {\bibinfo {author} {\bibfnamefont {A.}~\bibnamefont {Coladangelo}}, \bibinfo {author} {\bibfnamefont {A.}~\bibnamefont {Grilo}}, \bibinfo {author} {\bibfnamefont {S.}~\bibnamefont {Jeffery}},\ and\ \bibinfo {author} {\bibfnamefont {T.}~\bibnamefont {Vidick}},\ }\bibfield  {title} {\bibinfo {title} {Verifier-on-a-leash: new schemes for verifiable delegated quantum computation, with quasilinear resources},\ }\href {https://arxiv.org/abs/1708.07359} {\bibfield  {journal} {\bibinfo  {journal} {arXiv}\ } (\bibinfo {year} {2020})},\ \Eprint {https://arxiv.org/abs/1708.07359} {arXiv:1708.07359 [quant-ph]} \BibitemShut {NoStop}%
\bibitem [{\citenamefont {Zhang}\ \emph {et~al.}(2019{\natexlab{b}})\citenamefont {Zhang}, \citenamefont {Chen}, \citenamefont {Yin}, \citenamefont {Peng}, \citenamefont {Hu}, \citenamefont {Hou}, \citenamefont {Zhou}, \citenamefont {Yu}, \citenamefont {Ye}, \citenamefont {Zhou}, \citenamefont {Xu}, \citenamefont {Tang}, \citenamefont {Xu}, \citenamefont {Han}, \citenamefont {Liu}, \citenamefont {Li},\ and\ \citenamefont {Guo}}]{ZCPZ19}%
  \BibitemOpen
  \bibfield  {author} {\bibinfo {author} {\bibfnamefont {W.-H.}\ \bibnamefont {Zhang}}, \bibinfo {author} {\bibfnamefont {G.}~\bibnamefont {Chen}}, \bibinfo {author} {\bibfnamefont {P.}~\bibnamefont {Yin}}, \bibinfo {author} {\bibfnamefont {X.-X.}\ \bibnamefont {Peng}}, \bibinfo {author} {\bibfnamefont {X.-M.}\ \bibnamefont {Hu}}, \bibinfo {author} {\bibfnamefont {Z.-B.}\ \bibnamefont {Hou}}, \bibinfo {author} {\bibfnamefont {Z.-Y.}\ \bibnamefont {Zhou}}, \bibinfo {author} {\bibfnamefont {S.}~\bibnamefont {Yu}}, \bibinfo {author} {\bibfnamefont {X.-J.}\ \bibnamefont {Ye}}, \bibinfo {author} {\bibfnamefont {Z.-Q.}\ \bibnamefont {Zhou}}, \bibinfo {author} {\bibfnamefont {X.-Y.}\ \bibnamefont {Xu}}, \bibinfo {author} {\bibfnamefont {J.-S.}\ \bibnamefont {Tang}}, \bibinfo {author} {\bibfnamefont {J.-S.}\ \bibnamefont {Xu}}, \bibinfo {author} {\bibfnamefont {Y.-J.}\ \bibnamefont {Han}}, \bibinfo {author} {\bibfnamefont {B.-H.}\ \bibnamefont {Liu}}, \bibinfo {author} {\bibfnamefont {C.-F.}\ \bibnamefont {Li}},\
  and\ \bibinfo {author} {\bibfnamefont {G.-C.}\ \bibnamefont {Guo}},\ }\bibfield  {title} {\bibinfo {title} {Experimental demonstration of robust self-testing for bipartite entangled states},\ }\href {https://doi.org/10.1038/s41534-018-0120-0} {\bibfield  {journal} {\bibinfo  {journal} {Npj Quantum Inf.}\ }\textbf {\bibinfo {volume} {5}},\ \bibinfo {pages} {4} (\bibinfo {year} {2019}{\natexlab{b}})}\BibitemShut {NoStop}%
\bibitem [{\citenamefont {Klyachko}\ \emph {et~al.}(2008)\citenamefont {Klyachko}, \citenamefont {Can}, \citenamefont {Binicio\ifmmode~\breve{g}\else \u{g}\fi{}lu},\ and\ \citenamefont {Shumovsky}}]{KCBS08}%
  \BibitemOpen
  \bibfield  {author} {\bibinfo {author} {\bibfnamefont {A.~A.}\ \bibnamefont {Klyachko}}, \bibinfo {author} {\bibfnamefont {M.~A.}\ \bibnamefont {Can}}, \bibinfo {author} {\bibfnamefont {S.}~\bibnamefont {Binicio\ifmmode~\breve{g}\else \u{g}\fi{}lu}},\ and\ \bibinfo {author} {\bibfnamefont {A.~S.}\ \bibnamefont {Shumovsky}},\ }\bibfield  {title} {\bibinfo {title} {Simple test for hidden variables in spin-1 systems},\ }\href {https://doi.org/10.1103/PhysRevLett.101.020403} {\bibfield  {journal} {\bibinfo  {journal} {Phys. Rev. Lett.}\ }\textbf {\bibinfo {volume} {101}},\ \bibinfo {pages} {020403} (\bibinfo {year} {2008})}\BibitemShut {NoStop}%
\bibitem [{\citenamefont {Clauser}\ \emph {et~al.}(1969)\citenamefont {Clauser}, \citenamefont {Horne}, \citenamefont {Shimony},\ and\ \citenamefont {Holt}}]{CHSH69}%
  \BibitemOpen
  \bibfield  {author} {\bibinfo {author} {\bibfnamefont {J.~F.}\ \bibnamefont {Clauser}}, \bibinfo {author} {\bibfnamefont {M.~A.}\ \bibnamefont {Horne}}, \bibinfo {author} {\bibfnamefont {A.}~\bibnamefont {Shimony}},\ and\ \bibinfo {author} {\bibfnamefont {R.~A.}\ \bibnamefont {Holt}},\ }\bibfield  {title} {\bibinfo {title} {Proposed experiment to test local hidden-variable theories},\ }\href {https://doi.org/10.1103/PhysRevLett.23.880} {\bibfield  {journal} {\bibinfo  {journal} {Phys. Rev. Lett.}\ }\textbf {\bibinfo {volume} {23}},\ \bibinfo {pages} {880} (\bibinfo {year} {1969})}\BibitemShut {NoStop}%
\bibitem [{\citenamefont {Kochen}\ and\ \citenamefont {Specker}(1967)}]{KS67}%
  \BibitemOpen
  \bibfield  {author} {\bibinfo {author} {\bibfnamefont {S.}~\bibnamefont {Kochen}}\ and\ \bibinfo {author} {\bibfnamefont {E.~P.}\ \bibnamefont {Specker}},\ }\bibfield  {title} {\bibinfo {title} {The problem of hidden variables in quantum mechanics},\ }\href {http://www.jstor.org/stable/24902153} {\bibfield  {journal} {\bibinfo  {journal} {J. Math. Mech.}\ }\textbf {\bibinfo {volume} {17}},\ \bibinfo {pages} {59} (\bibinfo {year} {1967})}\BibitemShut {NoStop}%
\bibitem [{\citenamefont {Um}\ \emph {et~al.}(2020)\citenamefont {Um}, \citenamefont {Zhao}, \citenamefont {Zhang}, \citenamefont {Wang}, \citenamefont {Wang}, \citenamefont {Qiao}, \citenamefont {Zhou}, \citenamefont {Ma},\ and\ \citenamefont {Kim}}]{UZZK20}%
  \BibitemOpen
  \bibfield  {author} {\bibinfo {author} {\bibfnamefont {M.}~\bibnamefont {Um}}, \bibinfo {author} {\bibfnamefont {Q.}~\bibnamefont {Zhao}}, \bibinfo {author} {\bibfnamefont {J.}~\bibnamefont {Zhang}}, \bibinfo {author} {\bibfnamefont {P.}~\bibnamefont {Wang}}, \bibinfo {author} {\bibfnamefont {Y.}~\bibnamefont {Wang}}, \bibinfo {author} {\bibfnamefont {M.}~\bibnamefont {Qiao}}, \bibinfo {author} {\bibfnamefont {H.}~\bibnamefont {Zhou}}, \bibinfo {author} {\bibfnamefont {X.}~\bibnamefont {Ma}},\ and\ \bibinfo {author} {\bibfnamefont {K.}~\bibnamefont {Kim}},\ }\bibfield  {title} {\bibinfo {title} {Randomness expansion secured by quantum contextuality},\ }\href {https://doi.org/10.1103/PhysRevApplied.13.034077} {\bibfield  {journal} {\bibinfo  {journal} {Phys. Rev. Appl.}\ }\textbf {\bibinfo {volume} {13}},\ \bibinfo {pages} {034077} (\bibinfo {year} {2020})}\BibitemShut {NoStop}%
\bibitem [{\citenamefont {Singh}\ \emph {et~al.}(2017)\citenamefont {Singh}, \citenamefont {Bharti},\ and\ \citenamefont {Arvind}}]{JKA17}%
  \BibitemOpen
  \bibfield  {author} {\bibinfo {author} {\bibfnamefont {J.}~\bibnamefont {Singh}}, \bibinfo {author} {\bibfnamefont {K.}~\bibnamefont {Bharti}},\ and\ \bibinfo {author} {\bibnamefont {Arvind}},\ }\bibfield  {title} {\bibinfo {title} {Quantum key distribution protocol based on contextuality monogamy},\ }\href {https://doi.org/10.1103/PhysRevA.95.062333} {\bibfield  {journal} {\bibinfo  {journal} {Phys. Rev. A}\ }\textbf {\bibinfo {volume} {95}},\ \bibinfo {pages} {062333} (\bibinfo {year} {2017})}\BibitemShut {NoStop}%
\bibitem [{\citenamefont {Cabello}\ \emph {et~al.}(2014)\citenamefont {Cabello}, \citenamefont {Severini},\ and\ \citenamefont {Winter}}]{CSW14}%
  \BibitemOpen
  \bibfield  {author} {\bibinfo {author} {\bibfnamefont {A.}~\bibnamefont {Cabello}}, \bibinfo {author} {\bibfnamefont {S.}~\bibnamefont {Severini}},\ and\ \bibinfo {author} {\bibfnamefont {A.}~\bibnamefont {Winter}},\ }\bibfield  {title} {\bibinfo {title} {Graph-theoretic approach to quantum correlations},\ }\href {https://doi.org/10.1103/PhysRevLett.112.040401} {\bibfield  {journal} {\bibinfo  {journal} {Phys. Rev. Lett.}\ }\textbf {\bibinfo {volume} {112}},\ \bibinfo {pages} {040401} (\bibinfo {year} {2014})}\BibitemShut {NoStop}%
\bibitem [{\citenamefont {Singh}\ \emph {et~al.}(2023)\citenamefont {Singh}, \citenamefont {Singh}, \citenamefont {Dorai},\ and\ \citenamefont {Arvind}}]{DJKA23}%
  \BibitemOpen
  \bibfield  {author} {\bibinfo {author} {\bibfnamefont {D.}~\bibnamefont {Singh}}, \bibinfo {author} {\bibfnamefont {J.}~\bibnamefont {Singh}}, \bibinfo {author} {\bibfnamefont {K.}~\bibnamefont {Dorai}},\ and\ \bibinfo {author} {\bibnamefont {Arvind}},\ }\bibfield  {title} {\bibinfo {title} {Monogamy relations of entropic non-contextual inequalities and their experimental demonstration},\ }\href {https://doi.org/10.1209/0295-5075/acd954} {\bibfield  {journal} {\bibinfo  {journal} {Europhys. Lett.}\ }\textbf {\bibinfo {volume} {142}},\ \bibinfo {pages} {68001} (\bibinfo {year} {2023})}\BibitemShut {NoStop}%
\bibitem [{\citenamefont {Xu}\ \emph {et~al.}(2023)\citenamefont {Xu}, \citenamefont {Steinberg}, \citenamefont {Singh}, \citenamefont {L{\'{o}}pez-Tarrida}, \citenamefont {Portillo},\ and\ \citenamefont {Cabello}}]{ZJJAJA23}%
  \BibitemOpen
  \bibfield  {author} {\bibinfo {author} {\bibfnamefont {Z.-P.}\ \bibnamefont {Xu}}, \bibinfo {author} {\bibfnamefont {J.}~\bibnamefont {Steinberg}}, \bibinfo {author} {\bibfnamefont {J.}~\bibnamefont {Singh}}, \bibinfo {author} {\bibfnamefont {A.~J.}\ \bibnamefont {L{\'{o}}pez-Tarrida}}, \bibinfo {author} {\bibfnamefont {J.~R.}\ \bibnamefont {Portillo}},\ and\ \bibinfo {author} {\bibfnamefont {A.}~\bibnamefont {Cabello}},\ }\bibfield  {title} {\bibinfo {title} {Graph-theoretic approach to {B}ell experiments with low detection efficiency},\ }\href {https://doi.org/10.22331/q-2023-02-16-922} {\bibfield  {journal} {\bibinfo  {journal} {{Quantum}}\ }\textbf {\bibinfo {volume} {7}},\ \bibinfo {pages} {922} (\bibinfo {year} {2023})}\BibitemShut {NoStop}%
\bibitem [{\citenamefont {Singh}\ \emph {et~al.}(2021)\citenamefont {Singh}, \citenamefont {Ghosh}, \citenamefont {Goyal},\ and\ \citenamefont {Arvind}}]{JSS21}%
  \BibitemOpen
  \bibfield  {author} {\bibinfo {author} {\bibfnamefont {J.}~\bibnamefont {Singh}}, \bibinfo {author} {\bibfnamefont {S.}~\bibnamefont {Ghosh}}, \bibinfo {author} {\bibfnamefont {S.~K.}\ \bibnamefont {Goyal}},\ and\ \bibinfo {author} {\bibnamefont {Arvind}},\ }\bibfield  {title} {\bibinfo {title} {Role of {B}ell-{CHSH} violation and local filtering in quantum key distribution},\ }\href {https://doi.org/10.1016/j.physleta.2021.127158} {\bibfield  {journal} {\bibinfo  {journal} {Phys. Lett. A}\ }\textbf {\bibinfo {volume} {392}},\ \bibinfo {pages} {127158} (\bibinfo {year} {2021})}\BibitemShut {NoStop}%
\bibitem [{\citenamefont {Roy}\ \emph {et~al.}(2026)\citenamefont {Roy}, \citenamefont {Bera},\ and\ \citenamefont {Majumdar}}]{PSA26}%
  \BibitemOpen
  \bibfield  {author} {\bibinfo {author} {\bibfnamefont {P.}~\bibnamefont {Roy}}, \bibinfo {author} {\bibfnamefont {S.}~\bibnamefont {Bera}},\ and\ \bibinfo {author} {\bibfnamefont {A.}~\bibnamefont {Majumdar}},\ }\bibfield  {title} {\bibinfo {title} {Secure one-sided device-independent quantum key distribution under collective attacks with enhanced robustness},\ }\href {https://doi.org/10.1007/s11128-026-05073-9} {\bibfield  {journal} {\bibinfo  {journal} {Quantum Inf. Process.}\ }\textbf {\bibinfo {volume} {25}},\ \bibinfo {pages} {46} (\bibinfo {year} {2026})}\BibitemShut {NoStop}%
\bibitem [{\citenamefont {Lovasz}(1979)}]{L79}%
  \BibitemOpen
  \bibfield  {author} {\bibinfo {author} {\bibfnamefont {L.}~\bibnamefont {Lovasz}},\ }\bibfield  {title} {\bibinfo {title} {On the {S}hannon capacity of a graph},\ }\href {https://doi.org/10.1109/TIT.1979.1055985} {\bibfield  {journal} {\bibinfo  {journal} {IEEE Trans. Inf. Theory}\ }\textbf {\bibinfo {volume} {25}},\ \bibinfo {pages} {1} (\bibinfo {year} {1979})}\BibitemShut {NoStop}%
\bibitem [{\citenamefont {Alizadeh}\ \emph {et~al.}(1997)\citenamefont {Alizadeh}, \citenamefont {Haeberly},\ and\ \citenamefont {Overton}}]{AHO97}%
  \BibitemOpen
  \bibfield  {author} {\bibinfo {author} {\bibfnamefont {F.}~\bibnamefont {Alizadeh}}, \bibinfo {author} {\bibfnamefont {J.-P.~A.}\ \bibnamefont {Haeberly}},\ and\ \bibinfo {author} {\bibfnamefont {M.~L.}\ \bibnamefont {Overton}},\ }\bibfield  {title} {\bibinfo {title} {Complementarity and nondegeneracy in semidefinite programming},\ }\href {https://doi.org/10.1007/BF02614432} {\bibfield  {journal} {\bibinfo  {journal} {Math. Program.}\ }\textbf {\bibinfo {volume} {77}},\ \bibinfo {pages} {111} (\bibinfo {year} {1997})}\BibitemShut {NoStop}%
\bibitem [{\citenamefont {Ac{\'\i}n}\ \emph {et~al.}(2015)\citenamefont {Ac{\'\i}n}, \citenamefont {Fritz}, \citenamefont {Leverrier},\ and\ \citenamefont {Sainz}}]{AFLS15}%
  \BibitemOpen
  \bibfield  {author} {\bibinfo {author} {\bibfnamefont {A.}~\bibnamefont {Ac{\'\i}n}}, \bibinfo {author} {\bibfnamefont {T.}~\bibnamefont {Fritz}}, \bibinfo {author} {\bibfnamefont {A.}~\bibnamefont {Leverrier}},\ and\ \bibinfo {author} {\bibfnamefont {A.~B.}\ \bibnamefont {Sainz}},\ }\bibfield  {title} {\bibinfo {title} {A combinatorial approach to nonlocality and contextuality},\ }\href {https://doi.org/10.1007/s00220-014-2260-1} {\bibfield  {journal} {\bibinfo  {journal} {Commun. Math. Phys.}\ }\textbf {\bibinfo {volume} {334}},\ \bibinfo {pages} {533} (\bibinfo {year} {2015})}\BibitemShut {NoStop}%
\bibitem [{\citenamefont {Hammack}\ \emph {et~al.}(2011)\citenamefont {Hammack}, \citenamefont {Imrich}, \citenamefont {Klav{\v{z}}ar}, \citenamefont {Imrich},\ and\ \citenamefont {Klav{\v{z}}ar}}]{H11}%
  \BibitemOpen
  \bibfield  {author} {\bibinfo {author} {\bibfnamefont {R.~H.}\ \bibnamefont {Hammack}}, \bibinfo {author} {\bibfnamefont {W.}~\bibnamefont {Imrich}}, \bibinfo {author} {\bibfnamefont {S.}~\bibnamefont {Klav{\v{z}}ar}}, \bibinfo {author} {\bibfnamefont {W.}~\bibnamefont {Imrich}},\ and\ \bibinfo {author} {\bibfnamefont {S.}~\bibnamefont {Klav{\v{z}}ar}},\ }\href@noop {} {\emph {\bibinfo {title} {Handbook of product graphs}}},\ Vol.~\bibinfo {volume} {2}\ (\bibinfo  {publisher} {CRC press},\ \bibinfo {year} {2011})\BibitemShut {NoStop}%
\bibitem [{\citenamefont {Ac{\'\i}n}\ \emph {et~al.}(2017)\citenamefont {Ac{\'\i}n}, \citenamefont {Duan}, \citenamefont {Roberson}, \citenamefont {Sainz},\ and\ \citenamefont {Winter}}]{ADRSW17}%
  \BibitemOpen
  \bibfield  {author} {\bibinfo {author} {\bibfnamefont {A.}~\bibnamefont {Ac{\'\i}n}}, \bibinfo {author} {\bibfnamefont {R.}~\bibnamefont {Duan}}, \bibinfo {author} {\bibfnamefont {D.~E.}\ \bibnamefont {Roberson}}, \bibinfo {author} {\bibfnamefont {A.~B.}\ \bibnamefont {Sainz}},\ and\ \bibinfo {author} {\bibfnamefont {A.}~\bibnamefont {Winter}},\ }\bibfield  {title} {\bibinfo {title} {A new property of the {L}ov{\'a}sz number and duality relations between graph parameters},\ }\href {https://doi.org/10.1016/j.dam.2016.04.028} {\bibfield  {journal} {\bibinfo  {journal} {Discrete Appl. Math.}\ }\textbf {\bibinfo {volume} {216}},\ \bibinfo {pages} {489} (\bibinfo {year} {2017})}\BibitemShut {NoStop}%
\bibitem [{\citenamefont {Bharti}\ \emph {et~al.}(2019{\natexlab{b}})\citenamefont {Bharti}, \citenamefont {Ray}, \citenamefont {Varvitsiotis}, \citenamefont {Cabello},\ and\ \citenamefont {Kwek}}]{KMAAL19}%
  \BibitemOpen
  \bibfield  {author} {\bibinfo {author} {\bibfnamefont {K.}~\bibnamefont {Bharti}}, \bibinfo {author} {\bibfnamefont {M.}~\bibnamefont {Ray}}, \bibinfo {author} {\bibfnamefont {A.}~\bibnamefont {Varvitsiotis}}, \bibinfo {author} {\bibfnamefont {A.}~\bibnamefont {Cabello}},\ and\ \bibinfo {author} {\bibfnamefont {L.-C.}\ \bibnamefont {Kwek}},\ }\bibfield  {title} {\bibinfo {title} {Local certification of programmable quantum devices of arbitrary high dimensionality},\ }\bibfield  {journal} {\bibinfo  {journal} {arXiv}\ }\href {https://doi.org/10.48550/arXiv.1911.09448} {10.48550/arXiv.1911.09448} (\bibinfo {year} {2019}{\natexlab{b}}),\ \Eprint {https://arxiv.org/abs/1911.09448} {arXiv:1911.09448 [quant-ph]} \BibitemShut {NoStop}%
\bibitem [{\citenamefont {{MOSEK ApS}}(2025)}]{mosek25}%
  \BibitemOpen
  \bibfield  {author} {\bibinfo {author} {\bibnamefont {{MOSEK ApS}}},\ }\href {https://docs.mosek.com/latest/toolbox/index.html} {\emph {\bibinfo {title} {The {MOSEK} Optimization Toolbox for {MATLAB} Manual. Version 11.0}}} (\bibinfo {year} {2025})\BibitemShut {NoStop}%
\bibitem [{\citenamefont {Ghosh}(2026)}]{GBS26}%
  \BibitemOpen
  \bibfield  {author} {\bibinfo {author} {\bibfnamefont {D.}~\bibnamefont {Ghosh}},\ }\href@noop {} {\bibinfo {title} {{Numerical verification of uniqueness of the primal SDP solution}}},\ \bibinfo {howpublished} {\url{https://github.com/Dhananjoy2001/Parallel_self_testing}} (\bibinfo {year} {2026})\BibitemShut {NoStop}%
\bibitem [{\citenamefont {Boyd}\ and\ \citenamefont {Vandenberghe}(2004)}]{BV04}%
  \BibitemOpen
  \bibfield  {author} {\bibinfo {author} {\bibfnamefont {S.}~\bibnamefont {Boyd}}\ and\ \bibinfo {author} {\bibfnamefont {L.}~\bibnamefont {Vandenberghe}},\ }\href@noop {} {\emph {\bibinfo {title} {Convex optimization}}}\ (\bibinfo  {publisher} {{C}ambridge {U}niversity {P}ress},\ \bibinfo {year} {2004})\BibitemShut {NoStop}%
\bibitem [{\citenamefont {Das}\ \emph {et~al.}(2018)\citenamefont {Das}, \citenamefont {Gutman}, \citenamefont {Milovanović}, \citenamefont {Milovanović},\ and\ \citenamefont {Furtula}}]{DAS18}%
  \BibitemOpen
  \bibfield  {author} {\bibinfo {author} {\bibfnamefont {K.~C.}\ \bibnamefont {Das}}, \bibinfo {author} {\bibfnamefont {I.}~\bibnamefont {Gutman}}, \bibinfo {author} {\bibfnamefont {I.}~\bibnamefont {Milovanović}}, \bibinfo {author} {\bibfnamefont {E.}~\bibnamefont {Milovanović}},\ and\ \bibinfo {author} {\bibfnamefont {B.}~\bibnamefont {Furtula}},\ }\bibfield  {title} {\bibinfo {title} {Degree-based energies of graphs},\ }\href {https://doi.org/https://doi.org/10.1016/j.laa.2018.05.027} {\bibfield  {journal} {\bibinfo  {journal} {Linear Algebra Appl.}\ }\textbf {\bibinfo {volume} {554}},\ \bibinfo {pages} {185} (\bibinfo {year} {2018})}\BibitemShut {NoStop}%
\bibitem [{\citenamefont {Zeng}\ \emph {et~al.}(2025)\citenamefont {Zeng}, \citenamefont {Zheng}, \citenamefont {Li}, \citenamefont {He},\ and\ \citenamefont {Li}}]{XXCMX25}%
  \BibitemOpen
  \bibfield  {author} {\bibinfo {author} {\bibfnamefont {X.-N.}\ \bibnamefont {Zeng}}, \bibinfo {author} {\bibfnamefont {X.-W.}\ \bibnamefont {Zheng}}, \bibinfo {author} {\bibfnamefont {C.}~\bibnamefont {Li}}, \bibinfo {author} {\bibfnamefont {M.-F.}\ \bibnamefont {He}},\ and\ \bibinfo {author} {\bibfnamefont {X.}~\bibnamefont {Li}},\ }\bibfield  {title} {\bibinfo {title} {Weighted strong product graphs based on frequency equilibrium},\ }\href {https://api.semanticscholar.org/CorpusID:283923137} {\bibfield  {journal} {\bibinfo  {journal} {2025 International Conference on Machine Learning and Cybernetics (ICMLC)}\ ,\ \bibinfo {pages} {93}} (\bibinfo {year} {2025})}\BibitemShut {NoStop}%
\bibitem [{\citenamefont {Axler}(2024)}]{S24}%
  \BibitemOpen
  \bibfield  {author} {\bibinfo {author} {\bibfnamefont {S.}~\bibnamefont {Axler}},\ }\href@noop {} {\emph {\bibinfo {title} {Linear algebra done right}}}\ (\bibinfo  {publisher} {Springer},\ \bibinfo {year} {2024})\BibitemShut {NoStop}%
\end{thebibliography}

%

\newpage

\appendix

\onecolumngrid
\setcounter{theorem}{0}
\setcounter{definition}{0}
\setcounter{lemma}{0}

\section{Proofs of main Theorems and Lemmas}
\label{appen:proofs}

In this section, we prove the theorems stated in the main draft. We also prove a few lemmas which are required for the proof of the theorems as well.

\begin{theorem}\label{thm:1}
     Consider a set of $n$ NC inequalities $\lbrace \beta_k \rbrace_k$ associated with exclusivity graphs $\lbrace \mathcal{G}_k \rbrace_k$ respectively, such that each NC inequality $\beta_k$ gives a $(\epsilon_k,\frac{1}{2})$-robust self-test for an ideal quantum realization $\mathcal{S}_k=\{\rho_k = \ket{v_{0_k}}\!\bra{v_{0_k}},\{\Pi_{i_k} = \ket{v_{i_k}}\!\bra{v_{i_k}}\}_{i_k=1}^{N_k}\}$. Then, one can define a composite NC inequality $\beta$ associated with a composite graph $\mathcal{G} = \bigvee_{k=1}^{n}\mathcal{G}_k$. If the composite ideal quantum realization $\mathcal{S} = \lbrace \rho = \bigotimes_{k = 1}^{n}\ket{v_{0_k}}\!\bra{v_{0_k}}, \lbrace \Pi_i = \bigotimes_{k = 1}^{n}\ket{v_{i_k}}\!\bra{v_{i_k}} \rbrace_{i}\rbrace$ achieves $\beta = \beta_{\text{QT}}$ such that $\beta_{\text{QT}}>\beta_{\text{NCHV}}$, then the NC inequality $\beta$ provides a $(\epsilon,\frac{1}{2})$-robust self-test for the quantum realization $\mathcal{S}$. Explicitly, for every implemented quantum strategy $\tilde{\mathcal{S}} = \left\{\tilde{\rho} = \ket{\tilde{v}_0}\!\bra{\tilde{v}_0},\{\tilde{\Pi}_i = \ket{\tilde{v}_i}\!\bra{\tilde{v}_i}\}_{i}\right\}$ that attains $\beta=\beta_{\text{QT}}-\epsilon$ then, there exists an isometry $V$ such that
    \begin{equation}
\norm{V\ket{v_i}\!\bra{v_i}V^\dagger-\ket{\tilde{v_i}}\bra{\tilde{v_i}}}\leq \mathcal{O}(\sqrt{\epsilon}), \quad 0\leq i\leq N.
        \label{eq:parallel_self_test_robust}
    \end{equation}
\end{theorem}

\begin{proof}

The proof can be broken into three main steps. In the first step we remark on an earlier result~\cite{KBMA19} which relates self-testing of quantum realizations with graph-theoretic quantities using semi-definite programming (SDP) techniques~\cite{AHO97,BV04}. The crucial takeaway is that an NC inequality self-tests the ideal quantum realization if and only if the corresponding primal Lov\'asz theta SDP admits a unique optimal solution. We utilize this key take-away from the results of Ref.~\cite{KBMA19} in the steps two and three of the proof. Specifically, in the second step, we show that the optimal solution of the primal SDP associated with the composite NC inequality factorizes into the optimal solutions of the individual NC inequalities. In the third step, we utilize this property to show that the primal optimal solution of the composite NC inequality is unique whenever the individual NC inequalities admit self-tests, thereby proving self-testing from $n$-parallel NC inequalities.

We begin with the first step and note that the maximum quantum value $\beta^{(k)}_{\mathrm{QT}}$ of an NC inequality 
\begin{equation}
    \beta_k := \sum_{i_k = 1}^{N_k}p(1|i_k) \leq \beta_{\text{NCHV}}^{(k)},
    \label{eq:NC_each_graph}
\end{equation}
 associated with an exclusivity graph $\mathcal{G}_k$ is given by the SDP
    \begin{equation}
    \begin{aligned}
        \vartheta(\mathcal{G}_k)&=\max\sum_{i=1}^{N_k} X^{(k)}_{ii} \\
     \text{s.t.}\quad &X^{(k)}_{00}=1,\quad\quad\quad~~ X^{(k)}\in \mathcal{S}_+^{1+N_k}\\
     &X^{(k)}_{i0}=X^{(k)}_{ii},\quad\quad1\leq i\leq N_k,\\
    & X^{(k)}_{i j}=0,\quad\quad\quad~~ v_{i}\sim v_{j},
    \label{eq:SDP_for_k}
    \end{aligned}
\end{equation}
where $\mathcal{S}_+^{1+N_k}$ is the cone of positive semi-definite matrices of dimensions $N_k+1$, and $X^{(k)}$ is the Gram matrix constructed by the quantum realization $\mathcal{S}_k$.

As shown in Ref.~\cite{KBMA19}, the optimal solution to Eq.~\eqref{eq:SDP_for_k} is given by
\begin{equation} 
X_k^*= \left( \begin{tabular}{c|c} 
$1$ & $x_{k}^{\intercal}$\\
\hline $x_{k}$ & $\Bar{X}_{k}$
\end{tabular} \right), 
\label{eq:primal_optimal_k} 
\end{equation}
where $x_k$ is a column vector in $\mathbb{R}^{N_k}$ and $\Bar{X}_k$ is a real square matrix of size $N_k \times N_k$. Most importantly, we note that $\Bar{X}_k$ can be written as
\begin{equation}
\Bar{X}_k= \mathcal{I}_k+A'_{\Bar{\mathcal{G}}_k},
\label{eq:barX_k}
\end{equation}
where $ \mathcal{I}_k$ denotes a diagonal matrix of dimension $N_k\times N_k$ whose diagonal entries represent the optimal probability (treated as vertex weights) associated with each vertex of the complement graph $\bar{\mathcal{G}}_k$ such that
\begin{equation}
    (\mathcal{I}_k)_{ii}=p^{(k)}_{i} ~\forall i \in \lbrace 1,2,\ldots, N_k \rbrace,
    \label{eq:Ik}
\end{equation}
with
\begin{equation}
    \sum_{i=1}^{N_k} (\mathcal{I}_k)_{ii} = \vartheta(\mathcal{G}_k), 
    \label{eq:theta_k}
\end{equation}
and $A'_{\Bar{\mathcal{G}}_k}$ is the weighted adjacency matrix of the complement graph $\bar{\mathcal{G}}_k$~\cite{DAS18}. For a weighted graph $\bar{\mathcal{G}}_k=\{\mathcal{V}'_k,\mathcal{E}'_k\}$, the weighted adjacency matrix $A'_{\bar{\mathcal{G}}_k}$ is defined by
\begin{equation}
    (A'_{\bar{\mathcal{G}}_k})_{i j}=
\begin{cases}
\Lambda^{(k)}_{i j}, & (v_{i}, v_{j}) \in \mathcal{E}'_k,\\
0, & \text{otherwise},
\end{cases}
\end{equation}
where, $\Lambda^{(k)}_{i j}$ denotes the weight associated with the pair of adjacent vertices $(v_{i},v_{j})$. For the unweighted case $\Lambda^{(k)}_{ij}=1$. Here, the weight associated with the pair of adjacent vertices $(v_i,v_j)$ in the complement graph $\bar{\mathcal{G}}_k$ is given by the optimal value of corresponding SDP variable.

Following Ref.~\cite{KBMA19}, an NC inequality admits a self-test for the quantum realization $\mathcal{S}_k$ if and only if the primal optimal solution of the corresponding Lov\'asz theta SDP is unique. Following our theorem statement, we begin with the premise that each NC inequality $\beta_k$ admits a self-test for the quantum realization $\mathcal{S}_k$. This implies that Eq.~\eqref{eq:primal_optimal_k} is unique for all $k$.

With the premise established above, we now move to the composite NC inequality 
\begin{equation}
    \beta := \sum_{i=1}^N p(1|i)\leq \beta_{\text{NCHV}}.
    \label{eq:NC_inequality}
\end{equation}   
The maximum quantum value $\beta_{\text{QT}}$ of the composite NC inequality $\beta$ is given by
\begin{equation}
    \begin{aligned}
        \vartheta(\mathcal{G})&=\max\sum_{i=1}^{N} X_{ii} \\
     \text{s.t.}\quad &X_{00}=1,\quad\quad\quad~~ X\in \mathcal{S}_+^{1+N}\\
     &X_{i0}=X_{ii},\quad\quad1\leq i\leq N,\\
    & X_{i j}=0,\quad\quad\quad~~ v_{i}\sim v_{j},
    \label{eq:SDP_for_composite}
    \end{aligned}
\end{equation}
where $X$ is the Gram matrix constructed by some quantum realization $\mathcal{S}$. The optimal solution of this SDP is denoted by $X^*$, and any other feasible solution $\tilde{X}$ is said to be $\epsilon$-suboptimal if it satisfies $\sum_i \tilde{X}_{ii} \geq \vartheta(\mathcal{G})-\epsilon$.

We only need to show that for a composite NC inequality $\beta$, the primal optimal solution of the SDP~\eqref{eq:SDP_for_composite} is unique when the primal optimal solutions of the individual SDPs are unique.

Let a primal optimal solution be given by 
\begin{equation}
    X^*=
    \left(
    \begin{tabular}{c|c}
    $1$ & $x^{\intercal}$\\
         \hline
        $x$ & $\bar X$
    \end{tabular}
    \right),
    \label{eq:primal_optimal_parallel}
\end{equation}
where $x$ is a column vector in $\mathbb{R}^{N}$ and $\Bar{X}$ is a real square matrix of size $N \times N$. As earlier, we define $\Bar{X}$ as
\begin{equation}
\Bar{X}= \mathcal{I}+A'_{\bar{\mathcal{G}}},
\label{eq:barX}
\end{equation}
where $ \mathcal{I}$ denotes the diagonal matrix of dimension $N\times N$ whose diagonal entries represent the optimal probability (treated as vertex weight) associated with each vertex of the complement graph $\bar{\mathcal{G}}$ such that
\begin{equation}
   (\mathcal{I})_{ii} = p_i \quad \forall i \in \{1,2,\ldots, N\},
\end{equation} 
with
\begin{equation}
    \sum_{i=1}^{N} (\mathcal{I})_{ii} = \vartheta(\mathcal{G}), 
    \label{eq:theta}
\end{equation}
and $A'_{\Bar{\mathcal{G}}}$ is the weighted adjacency matrix of the complement graph $\bar{\mathcal{G}}$.

Moving on to the second step, we proceed to show that a primal optimal solution of the composite SDP can be decomposed as
\begin{equation}
X^*=\left( \begin{tabular}{c|c}
         $1$&$\bigotimes_{k = 1}^{n}x_k^{\intercal}$\\
         \hline
        $\bigotimes_{k = 1}^{n}x_k$ & $\bigotimes_{k = 1}^{n} \bar X_k$
    \end{tabular}\right).
    \label{eq:primal_comp_decom}
\end{equation}

We begin by first considering $n = 2$ NC inequalities in parallel denoted by $\beta_1$ and $\beta_2$ associated with the exclusivity graphs 
$\mathcal{G}_{1}$ and $\mathcal{G}_2$ respectively. For non-robust self-testing we also consider that each of the inequalities attains its maximum quantum value given by $\beta^{(1)}_{\rm QT}$ and $\beta^{(2)}_{\rm QT}$ with the primal optimal solutions denoted by $X^*_1$ and $X^*_2$ respectively.
The composite exclusivity graph of the two parallel NC inequalities is
given by $\mathcal{G}=\mathcal{G}_1\vee\mathcal{G}_2$. Using the relation between the OR and strong products of graphs, we have
\begin{equation}
    \bar{\mathcal{G}}
    =
    \overline{\mathcal{G}_1\vee\mathcal{G}_2}
    =
    \bar{\mathcal{G}}_1\boxtimes
    \bar{\mathcal{G}}_2.
\end{equation}
Following the weighted strong-product construction~\cite{XXCMX25,ADRSW17}, for a given assignment of weights, the weighted adjacency matrix of the product graph is given by
\begin{equation}
    A'_{\bar{\mathcal{G}}}
    =
    A'_{\bar{\mathcal{G}}_1}\otimes \mathcal{I}_2
    +
    \mathcal{I}_1\otimes A'_{\bar{\mathcal{G}}_2}
    +
    A'_{\bar{\mathcal{G}}_1}\otimes
    A'_{\bar{\mathcal{G}}_2}.
    \label{eq:comp_A_decom_strong}
\end{equation}

Next, we note that, following Ref.~\cite{ADRSW17}, the weight assigned to a vertex in the composite graph is given by the product of the weights assigned to the corresponding vertices in the individual graphs. Therefore, if the composite vertex $v_i$ corresponds to the vertex pair $(v_{i_1},v_{i_2})$, then
\begin{equation}
    p_i = p^{(1)}_{i_1} p^{(2)}_{i_2},
    \label{eq:fact_pi}
\end{equation}
where $p^{(1)}_{i_1}$ and $p^{(2)}_{i_2}$ denote the weights associated with the vertices $v_{i_1}$ and $v_{i_2}$ in $\mathcal{G}_1$ and $\mathcal{G}_2$, respectively. Here, it should be noted that this factorization alone does not imply self-testing. Self-testing additionally requires the corresponding optimal solution of the composite Lov\'asz theta SDP to be unique.

Consequently, using Eq.~\eqref{eq:fact_pi}, the diagonal weight matrix of the composite graph factorizes as the tensor product of the diagonal weight matrices of the individual graphs, i.e.,
\begin{equation}
    \mathcal{I}=\mathcal{I}_1 \otimes \mathcal{I}_2.
    \label{eq:decom_I}
\end{equation}

Furthermore, from Eqs.~\eqref{eq:theta_k},~\eqref{eq:theta}, and~\eqref{eq:decom_I}, it follows that
\begin{equation}
    \vartheta(\mathcal{G}) = \vartheta(\mathcal{G}_1) \vartheta(\mathcal{G}_2).
\end{equation} 

Substituting Eqs.~\eqref{eq:comp_A_decom_strong} and~\eqref{eq:decom_I} into Eq.~\eqref{eq:barX},
we obtain
\begin{equation}
    \begin{aligned}
        \bar{X}
        &=
        \mathcal{I}+A'_{\bar{\mathcal{G}}}\\
        &=
        \mathcal{I}_1\otimes\mathcal{I}_2
        +A'_{\bar{\mathcal{G}}_1}\otimes\mathcal{I}_2
        +\mathcal{I}_1\otimes A'_{\bar{\mathcal{G}}_2}
        +A'_{\bar{\mathcal{G}}_1}\otimes
        A'_{\bar{\mathcal{G}}_2}\\
        &=
        \left(
        \mathcal{I}_1+A'_{\bar{\mathcal{G}}_1}
        \right)
        \otimes
        \left(
        \mathcal{I}_2+A'_{\bar{\mathcal{G}}_2}
        \right).
    \end{aligned}
    \label{eq:decom_I_CA}
\end{equation}

Therefore, from Eqs.~\eqref{eq:barX_k}, ~\eqref{eq:barX} and~\eqref{eq:decom_I_CA} we can conclude that 
\begin{equation}
    \bar X = \bar X_1 \otimes \bar X_2.
    \label{eq:decom_Xbar}
\end{equation}

We emphasize that Eqs.~\eqref{eq:comp_A_decom_strong},~\eqref{eq:fact_pi} and~\eqref{eq:decom_I_CA} follow from the weighted strong-product construction together with the multiplicative rule for the vertex weights. Once the optimal weights of the individual complement graphs are fixed, the corresponding weights and weighted adjacency matrix of the composite graph are fixed. Consequently, for every primal optimal solution of the composite SDP, the block $\bar X$ takes the form of Eq.~\eqref{eq:decom_Xbar}.

Next, we establish relation between $x_i$ and $x$ in Eqs.~\eqref{eq:primal_optimal_k} and~\eqref{eq:primal_optimal_parallel} respectively. We note that the diagonal entries of $\bar X$ can be written as
\begin{equation}
\begin{aligned}
      (\bar X)_{ii} &= (\bar X_1)_{i_1i_1} (\bar X_2)_{i_2i_2},\\
     \implies\quad x_i &= (x_1)_{i_1} (x_2)_{i_2},\\
      \implies\quad x &= x_1 \otimes x_2,
\end{aligned}
\end{equation}
where the second equality follows from the constraints of Eqs.~\eqref{eq:SDP_for_k} and~\eqref{eq:SDP_for_composite} respectively.

By mathematical induction, the above result extends to $n$ parallel NC inequalities, so that a primal optimal solution for the composite scenario can be written as in Eq.~\eqref{eq:primal_comp_decom}.

We also prove that if all the primal optimal solutions $X^*_k$ are positive semi-definite, then 
$X^*$ is also positive semi-definite in Lemma~\ref{lemma:PDS_combine_dual_optimal_solution}.

In the third step, we show that the primal optimal solution of the composite SDP in Eq.~\eqref{eq:primal_comp_decom} is unique whenever the primal optimal solutions of all individual SDPs are unique. By the given condition of the theorem, each individual NC inequality provides a self-test of the corresponding quantum realization. Therefore, the associated primal optimal solution $X_k^*$ is unique for every $k$. Consequently, all blocks appearing in $X_k^*$ are uniquely determined.

From Eq.~\eqref{eq:primal_comp_decom}, the composite primal optimal solution $X^*$ is completely determined by  $x_k$ and the matrices $\bar{X}_k$. Since $x_k$ and $\bar{X}_k$ are uniquely determined for every $k$, the decomposition in Eq.~\eqref{eq:primal_comp_decom} uniquely specifies every block of $X^*$. Consequently, $X^*$ is unique, and hence the primal optimal solution of the composite SDP is unique.

Therefore, the composite NC inequality $\beta$ provides a self-test for the composite realization $\mathcal{S}$ whenever each individual NC inequality provides a self-test for the corresponding individual realization.

We now quantify the robustness of the self-testing statement for the composite realization. If $X^*$ be the optimal solution of the SDP in Eq.~\eqref{eq:SDP_for_composite}, and $\tilde{X}$ be any $\epsilon$-suboptimal solution of the same SDP then, it follows from Ref.~\cite{KBMA19} that
\begin{equation}
\norm{\tilde{X} - X^*}_F \leq \mathcal{O}(\epsilon),
\end{equation}
where $\norm{\cdot}_F$ denotes the Frobenius norm~\cite{S24}. Moreover, since $\tilde{X}$ and $X^*$ are positive semidefinite matrices, their closeness in Frobenius norm implies that the corresponding vectors in their Gram decompositions are $\mathcal{O}(\sqrt{\epsilon})$-close in $2$-norm, thereby guaranteeing that the implemented realization is also $\mathcal{O}(\sqrt{\epsilon})$-close to the ideal realization (up to global isometries).

\end{proof}


\begin{theorem}\label{thm:2}
   Consider an NC inequality $\beta$ associated with an exclusivity graph $\mathcal{G}$ that provides a $(\epsilon,\frac{1}{2})$-robust self-test for an ideal quantum realization $\mathcal{S}=\{\rho,\{\Pi_i\}_{i=1}^N\}$. If $\mathcal{G} = \bigvee_{k=1}^{n}\mathcal{G}_k$, then every individual NC inequality $\beta_k$ corresponding to the exclusivity graph $\mathcal{G}_k$ provides a $(\epsilon_k,\frac{1}{2})$-robust self-test for the corresponding quantum realization $\mathcal{S}_k=\{\rho_k,\{\Pi_{i_k}\}_{i_k}\}$, provided that $\mathcal{S}_k$ achieves $\beta_k=\beta_{\text{QT}}^{(k)}$, such that $\beta_{\text{QT}}^{(k)} >\beta_{\text{NCHV}}^{(k)}$. Explicitly, for every implemented quantum strategy $\tilde{S}_k = \{\tilde{\rho}_k,\{\tilde{\Pi}_{i_k}\}_{i_k}\}$ that attains $\beta_k=\beta_{\text{QT}}^{(k)}-\epsilon_k$, there will exist an isometry $V_k$ where
\begin{equation}
\norm{V_k\ket{v_{i_k}}\!\bra{v_{i_k}}V_k^\dagger-\ket{\tilde{v}_{i_k}}\bra{\tilde{v}_{i_k}}}\leq \mathcal{O}(\sqrt{\epsilon_k}), \quad \forall~k,
\end{equation}
where $0\leq i_k\leq N_k$.
\label{eq:individual_self_test_robust}
\end{theorem}

\begin{proof}

For the proof, we follow the same approach as in the proof of Theorem~\ref{thm:1}. It is sufficient to prove the result for $n=2$, since the general case follows by induction. Let $\mathcal{G}=\mathcal{G}_1\vee\mathcal{G}_2$, and let
\begin{equation}
    X^*=
\left(
\begin{array}{c|c}
1 & x^{\intercal}\\
\hline
x & \bar X
\end{array}
\right)
\end{equation}
be the unique primal optimal solution of the Lov\'asz theta SDP corresponding to the composite graph $\mathcal{G}$. From Eq.~\eqref{eq:primal_comp_decom}, every primal optimal solution of the composite SDP admits the decomposition
\begin{equation}
X^*=
\left(
\begin{array}{c|c}
1 & x_1^{\intercal}\otimes x_2^{\intercal}\\
\hline
x_1\otimes x_2 & \bar X_1\otimes \bar X_2
\end{array}
\right),
\label{eq:proof_decomp}
\end{equation}
where
\begin{equation}
 X_k^*=
\left(
\begin{array}{c|c}
1 & x_k^{\intercal}\\
\hline
x_k & \bar X_k
\end{array}
\right),
\qquad \forall k\in\{1,2\},   
\end{equation}
are primal optimal solutions of the Lov\'asz theta SDPs corresponding to the graphs $\mathcal{G}_1$ and $\mathcal{G}_2$, respectively.

We now show that each individual SDP must also admit a unique primal optimal solution. Suppose, on the contrary, that the SDP corresponding to $\mathcal{G}_1$ does not admit a unique primal optimal solution. Then there exist at least two distinct optimal solutions, say $X_{1a}^*$ and $X_{1b}^*$, with $X_{1a}^*\neq X_{1b}^*$. Keeping the optimal solution $X_2^*$ fixed, Eq.~\eqref{eq:proof_decomp} yields two distinct primal optimal solutions of the composite SDP,
\begin{subequations}
\begin{equation}
    X_a^*=
\left(
\begin{array}{c|c}
1 & x_{1a}^{\intercal}\otimes x_2^{\intercal}\\
\hline
x_{1a}\otimes x_2 & \bar X_{1a}\otimes \bar X_2
\end{array}
\right),
\end{equation}
\begin{equation}
    X_b^*=
\left(
\begin{array}{c|c}
1 & x_{1b}^{\intercal}\otimes x_2^{\intercal}\\
\hline
x_{1b}\otimes x_2 & \bar X_{1b}\otimes \bar X_2
\end{array}
\right).
\end{equation} 
\end{subequations}

Since $X_{1a}^*\neq X_{1b}^*$, it follows that $X_a^*\neq X_b^*$. This contradicts the given condition that the composite SDP admits a unique primal optimal solution. Hence, the SDPs corresponding to $\mathcal{G}_1$ and $\mathcal{G}_2$ must each admit a unique primal optimal solution. Therefore, each individual NC inequality associated with $\mathcal{G}_1$ and $\mathcal{G}_2$ provides a self-test for the corresponding individual realization whenever the composite NC inequality provides the self-test for the composite quantum realization. By the principle of mathematical induction, this result extends naturally for any arbitrary $n$.

We now derive the robustness of the self-testing statement for an individual quantum realization. If $X_k^*$ be the optimal solution of the SDP in Eq.~\eqref{eq:SDP_for_k}, and $\tilde{X}_k$ be any $\epsilon_k$-suboptimal solution of the same SDP then, it follows from Ref.~\cite{KBMA19} that
\begin{equation}
\norm{\tilde{X}_k - X_k^*}_F \leq \mathcal{O}(\epsilon_k),
\end{equation}
where $\norm{\cdot}_F$ denotes the Frobenius norm. Moreover, since $\tilde{X}_k $ and $X_k^*$ are positive semidefinite matrices, their closeness in Frobenius norm implies that the corresponding vectors in their Gram decompositions are $\mathcal{O}(\sqrt{\epsilon_k})$-close in $2$-norm.
\end{proof}

\begin{lemma}\label{lemma:PDS_combine_dual_optimal_solution}
    Let $\lbrace\mathcal{G}_k\rbrace_k$ be a set of exclusivity graphs associated with a corresponding set of NC inequalities $\lbrace \beta_k\rbrace_k$. Denote by $\mathcal{G} = \bigvee_k \mathcal{G}_k$ the composite exclusivity graph corresponding to a composite NC inequality $\beta$. The primal optimal solution, $X^*$, of the SDP for the quantum maximum of the NC inequality $\beta$, as given in Eq.~\eqref{eq:primal_comp_decom}, is positive semi-definite if the primal optimal solutions corresponding to the individual NC inequalities $\beta_k$, is also positive semi-definite, that is $X^*_k \succeq 0~\forall k$.
\end{lemma}

\begin{proof}
To establish the proof, we make use of the Schur complement technique~\cite{BV04}.  Consider a symmetric matrix
\begin{equation}
    G=\begin{pmatrix}
A & B \\
B^{\intercal} & D
\end{pmatrix},
\end{equation}
where $A$ and $D$ are in general square matrices.  If $A$ is invertible, then the Schur complement of $A$ in $G$ is defined as
\begin{equation}
S = D - B^{\intercal} A^{-1} B.
\end{equation} 
The Schur complement provides a useful criterion for positive 
semi-definiteness. In particular, if $A$ is invertible, we have
\begin{equation}
G \succeq 0 \quad \Longleftrightarrow \quad A \succeq 0  \text{ and }  
D - B^{\intercal} A^{-1} B \succeq 0.
\end{equation}

Similarly, from Eq.~\eqref{eq:primal_comp_decom}, the Schur complement of $\vartheta(\mathcal{G})$ in $X^*$ is given by
\begin{equation}
    S=\bigotimes_{k=1}^n \bar X_{k}- xx^{\intercal}.
    \label{Eq.schur_complement_of_Z*}
\end{equation}
To prove that $X^*$ is positive semi-definite, it is sufficient to show that
\begin{subequations}
    \begin{equation}
        1\geq  0,
    \end{equation}
\begin{equation}
        \bigotimes_{k=1}^n \bar X_{k}
        -   xx^{\intercal} 
        \succeq 0
        \label{eq:schur_condition}
    \end{equation}
\end{subequations}

We only need to analyze the non-trivial condition in Eq.~\eqref{eq:schur_condition}. Using the relation $x=\bigotimes_{k=1}^nx_k$, we obtain  
\begin{equation}
        \begin{aligned}
             xx^{\intercal}&= \left(\bigotimes_{k=1}^nx_k\right)\left(\bigotimes_{k=1}^nx_k^{\intercal}\right)\\
            &= \bigotimes_{k=1}^n x_k x_k^{\intercal}
            \end{aligned}
            \label{Eq:second_part_schur_condition}
\end{equation}
On the other hand, since each realization is self-tested by its corresponding individual NC inequality then the primal optimal matrix $X_k^*$ associated with NC inequality $\beta_k$ must be positive semi-definite and by the Schur complement, we have
\begin{equation}
    \bar X_{k}\succeq x_{k}x_{k}^{\intercal} \quad \forall k.
    \label{Eq:schur_condition_Xk}
\end{equation}
Since the tensor product of positive semi-definite matrices is also positive semi-definite, it follows that
\begin{equation}
\bigotimes_{k=1}^n \bar X_k 
\succeq 
\bigotimes_{k=1}^n   x_k x_k^\intercal
=   xx^\intercal.
\end{equation}

Thus,
\begin{equation}
\bigotimes_{k=1}^n \bar X_k -   xx^\intercal \succeq 0.
\end{equation}

Therefore, both Schur complement conditions are satisfied, and we conclude that
\begin{equation}
X^* \succeq 0.
\end{equation}

By the principle of mathematical induction, this result extends naturally to the case of $n$ composite NC inequalities.

\end{proof}

\section{KCBS inequality and other self-testing examples}
\label{appsec:examples}

\subsection{Optimal quantum realization of the KCBS inequality}
\label{appsubsec:Parallel_KCBS}

\begin{figure}
    \centering
    \includegraphics[width=0.25\linewidth]{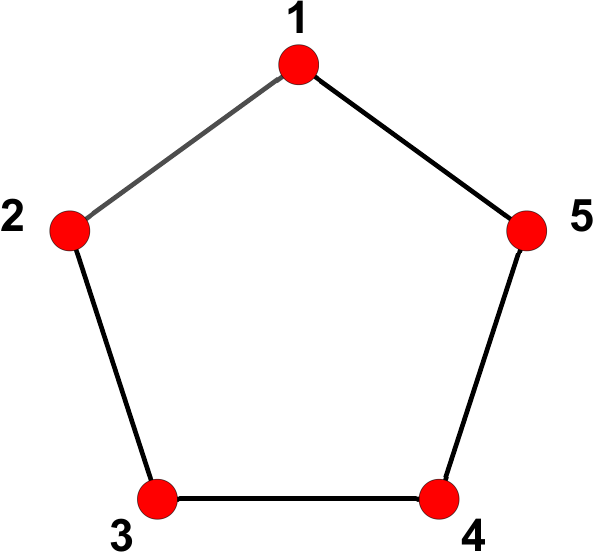}
    \caption{The exclusivity graph associated with KCBS inequality, denoted by $C_5$. The exclusivity graph has independence number $\alpha(C_5)=2$ and Lov\'asz number $\vartheta(C_5)=\sqrt{5}$.}
    \label{fig:ex_graph_KCBS}
\end{figure}

The KCBS inequality is the simplest NC inequality that demonstrates quantum contextuality. The KCBS inequality is given by 
\begin{equation}
    \beta = \sum_{i=1}^5 p(1|i) \leq 2,
    \label{eq:KCBS_inequality}
\end{equation}
where $p(1|i)$ is the probability of obtaining outcome $1$ for projector $\Pi_{i}$. The exclusivity graph associated with the KCBS inequality is the five-vertex regular cyclic graph (pentagon), denoted by $C_5$, as shown in Fig.~\ref{fig:ex_graph_KCBS}. The corresponding NC bound is given by the independence number, $\beta_{\text{NCHV}}=\alpha(C_5)=2$. Solving the Lovász theta SDP in Eq.~\eqref{eq:SDP_for_k} using the exclusivity relations of Fig.~\ref{fig:ex_graph_KCBS}, we obtain the maximum quantum value $\beta_{\text{QT}}=\vartheta(C_5)=\sqrt{5}$. This value is attained by the optimal qutrit realization
$\mathcal{S}=\{ \ket{v_{0}}\!\bra{v_{0}},\{ \ket{v_{i}}\!\bra{v_{i}}\}_{i=1}^{5}\}$, where
\begin{equation}
\begin{aligned}
    \ket{v_{0}} &= (1,0,0)^\intercal, \\
     \ket{v_{i}} &= (\cos(\theta),~ 
    \sin(\theta)\sin(\phi_{i}),~
    \sin(\theta)\cos(\phi_{i}))^\intercal,
\end{aligned}
\end{equation}
where $\cos^2(\theta)=[\cos(\pi/5)]/ [1+\cos(\pi/5)]$ and $\phi_{i}=4\pi i/5,$ for $1\leq i\leq 5$. 

\subsection{Self-testing from two parallel Bell-CHSH inequalities}
\label{appsubsec:Parallel_CHSH}

To illustrate another example for the applicability of Theorem~\ref{thm:1}, we consider two parallel Bell-CHSH inequalities, where a single instance of the Bell-CHSH inequality~\cite{CSW14,Bell64} is expressed as
\begin{equation}
\beta_j=\sum_{i_j=1}^8p(1|i_j)\leq 3, \qquad j\in \lbrace 1,2\rbrace
\label{eq:Bell_CHSH_inequality}
\end{equation}
where we have $\beta^{(j)}_{\text{NCHV}}=3$ and $\beta^{(j)}_{\text{QT}}=2+\sqrt{2}$ which can be achieved by a ququart (four-dimensional) quantum state and an ideal quantum realization $\mathcal{S}_j=\{ \ket{v_{0_j}}\!\bra{v_{0_j}},\{ \ket{v_{i_j}}\!\bra{v_{i_j}}\}_{i_j=1}^{8}\}$, where the vectors $\ket{v_{0_j}}$ and $\lbrace \ket{v_{i_j}}\rbrace_{i_j = 1}^{8}$ for $j = 1, 2$ are given as
\begin{equation}
\begin{aligned}
\ket{v_{0_j}}&=\left(\frac{1}{\sqrt{2}},0,0,\frac{1}{\sqrt{2}}\right)^\intercal\\
|v_{1_j}\rangle &= |A_{1,0}\rangle \otimes |B_{1,1}\rangle, \\
|v_{2_j}\rangle &= |A_{0,0}\rangle \otimes |B_{1,0}\rangle, \\
|v_{3_j}\rangle &= |A_{0,1}\rangle \otimes |B_{0,1}\rangle, \\
|v_{4_j}\rangle &= |A_{1,0}\rangle \otimes |B_{0,0}\rangle, \\
|v_{5_j}\rangle &= |A_{1,1}\rangle \otimes |B_{1,0}\rangle, \\
|v_{6_j}\rangle &= |A_{0,1}\rangle \otimes |B_{1,1}\rangle, \\
|v_{7_j}\rangle &= |A_{0,0}\rangle \otimes |B_{0,0}\rangle, \\
|v_{8_j}\rangle &= |A_{1,1}\rangle \otimes |B_{0,1}\rangle,
\end{aligned}
\label{eq:realization_CHSH}
\end{equation}
where,
\begin{equation}
\begin{aligned}
|A_{0,0}\rangle &= (1,0)^\intercal, \\
|A_{0,1}\rangle &= (0,-1)^\intercal,\\
|A_{1,0}\rangle &= (a,a)^\intercal, \\
|A_{1,1}\rangle &= (a,-a)^\intercal,\\
|B_{0,0}\rangle &= (c,d)^\intercal, \\
|B_{0,1}\rangle &= (d,-c)^\intercal,\\
|B_{1,0}\rangle &= (c,-d)^\intercal, \\
|B_{1,1}\rangle &= (-d,-c)^\intercal,
\end{aligned}
\end{equation}
with $a=1/\sqrt{2}$, $c=\cos(\pi/8)$, and $d=\sin(\pi/8)$.

\begin{figure}
    \centering
    \includegraphics[width=0.3\linewidth]{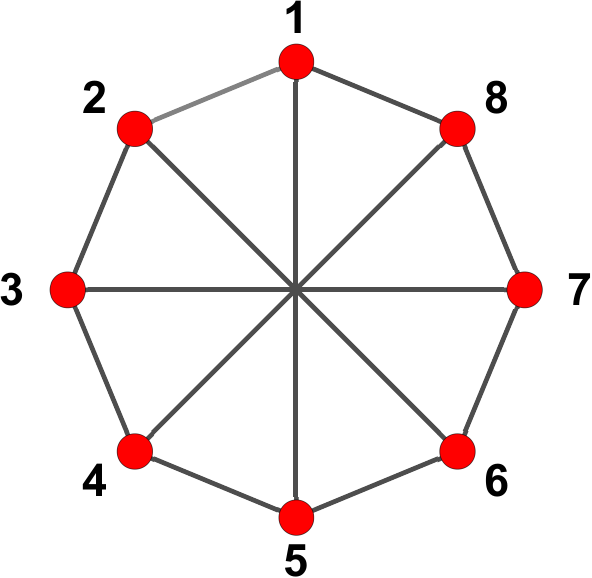}
    \caption{The exclusivity graph associated with Bell-CHSH inequality, denoted by $C_{i8}(1,4)$. The exclusivity graph has independence number $\alpha(C_{i8}(1,4))=2$ and Lov\'asz number $\vartheta(C_{i8}(1,4))=\sqrt{5}$.}
    \label{fig:ex_graph_Bell_CHSH}
\end{figure}

The exclusivity graph associated with the Bell-CHSH inequality is the eight-vertex circulant graph denoted by $C_{i_8}(1,4)$, as shown in Fig.~\ref{fig:ex_graph_Bell_CHSH}.

Following our framework, corresponding to the two parallel copies of Bell-CHSH inequalities, we first construct a composite inequality whose exclusivity graph is obtained as the OR graph product of the exclusivity graphs associated with the individual inequalities, i.e., $\mathcal{G}=C_{i_8}(1,4) \vee C_{i_8}(1,4)$. The resulting graph consists of $N = 64$ vertices and $1248$ edges. Moreover, we find that $\alpha(\mathcal{G})=9$, while $\vartheta(\mathcal{G})=(2+\sqrt{2})^2 \approx 11.65$, which can be achieved by the composite ququart realization $\mathcal{S} = \left\{\ket{v_0}\!\bra{v_0}, \left\{\ket{v_i}\!\bra{v_i}\right\}_{i=1}^{64}\right\}$,
where $ \ket{v_0}\!\bra{v_0} = \ket{v_{0_1}}\!\bra{v_{0_1}} \otimes \ket{v_{0_2}}\!\bra{v_{0_2}}$ and $\ket{v_i}\!\bra{v_i}=\ket{v_{i_1}}\!\bra{v_{i_1}} \otimes \ket{v_{i_2}}\!\bra{v_{i_2}}$. Since $\beta_{\text{QT}}>\beta_{\text{NCHV}}$, the conditions of Theorem~\ref{thm:1} are satisfied. 
Using the exclusivity relations of the composite graph, we construct the SDP given in Eq.~\eqref{eq:SDP_for_composite}, and numerically obtain a dual optimal solution $Z^*$.

Next, we construct the variable matrix $M$ subject to the constraints
\begin{equation}
    M_{00}=0,\quad M_{i0}=M_{ii},\quad
    M_{ij}=0 \;\; \forall\, v_i\sim v_j,
\label{eq:M}
\end{equation}
where the exclusivity relations $v_i\sim v_j$ are determined by the composite graph $\mathcal{G}=C_{i_8}(1,4)\vee C_{i_8}(1,4)$. We then solve the linear system $MZ^*=0$. As before, we reformulate the problem as a homogeneous linear system $Ax=0$ and analyze the corresponding coefficient matrix $A$. In the present case, we find that all singular values are significantly larger than the threshold value~\cite{GBS26}, which indicates that $A$ is numerically full rank, and hence the only admissible solution is $M=0$, which implies that the primal optimal solution of the SDP corresponding to the composite Bell-CHSH scenario is unique. It then follows from Theorem~\ref{thm:1} that the associated composite inequality self-tests the composite qutrit-ququart realization $\mathcal{S}$ with robustness $\mathcal{O}(\sqrt{\epsilon})$. We present a numerical package for the same in Ref.~\cite{GBS26}.

Here, as before, we consider the optimal quantum realization of the CHSH scenario to be a ququart realization rather than a two-qubit realization. Consequently, the equivalence between the ideal and implemented realizations is certified by a single global isometry acting on the entire composite ququart Hilbert space.

\subsection{Composite scenario with a non-self-testing NC inequality}
\label{appsubsec:Parallel_KCBS_C6}

\begin{figure}
    \centering
    \includegraphics[width=0.25\linewidth]{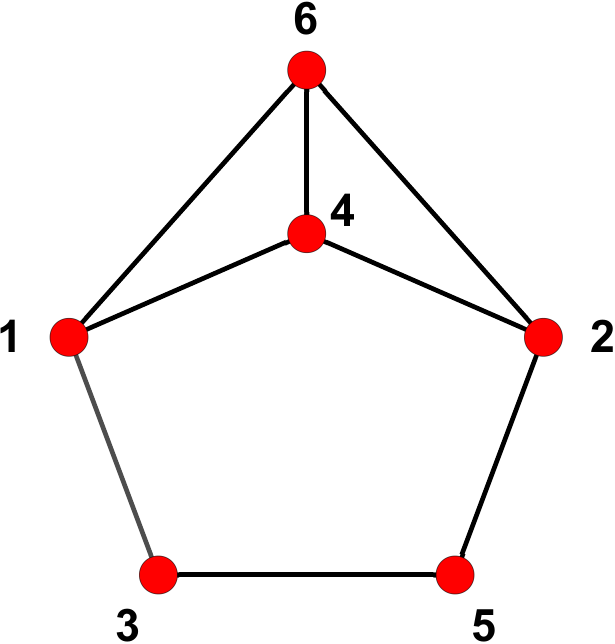}
    \caption{The exclusivity graph associated with NC inequality, given in Eq.~\eqref{eq:G6_inequality}, denoted by $\mathcal{G}_6$. The exclusivity graph has independence number $\alpha(\mathcal{G}_6)=2$ and Lov\'asz number $\vartheta(\mathcal{G}_6)=\sqrt{5}$.}
    \label{fig:Ex_graph_G6}
\end{figure}

\begin{figure}
    \centering
    \includegraphics[width=0.45\linewidth]{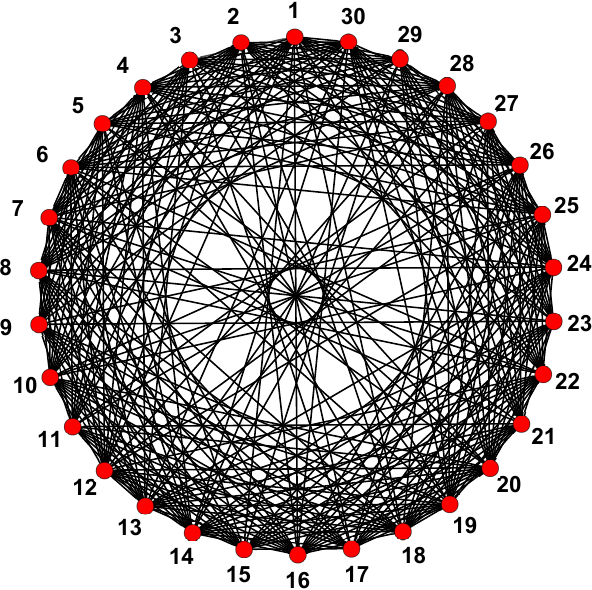}
    \caption{The exclusivity graph $\mathcal{G}=C_5 \vee \mathcal{G}_6$ corresponding to parallel tests of the KCBS inequality and the NC inequality given in Eq.~\eqref{eq:G6_inequality}. The resulting graph has $N=30$ vertices and $300$ edges, with $\alpha(\mathcal{G}) = 4$ and $\vartheta(\mathcal{G}) = 5$.}
    \label{fig:Ex_graph_parallel_KCBS_G6}
\end{figure}

In this section, we analyze a composite scenario consisting of one KCBS inequality and another NC inequality that which we term as $\mathcal{G}_6$ inequality. The latter inequality is interesting in the sense that even though it exhibits a quantum violation, it does not provide a self-test of the underlying optimal quantum realization. As such this example will serve to illustrate the case when one of the individual NC inequalities cannot admit a self-test, the composite NC inequality also does not.

The $\mathcal{G}_6$ NC inequality is given by
\begin{equation}
   \beta_3 = \sum_{i_3=1}^6 p(1|i_3) \leq 2,
    \label{eq:G6_inequality}
\end{equation}
and the exclusivity graph associated with this NC inequality is shown in Fig.~\ref{fig:Ex_graph_G6}. The corresponding NC bound is given by the independence number,
$\beta_{\text{NCHV}}=\alpha(\mathcal{G}_6)=2$. Solving the Lovász SDP in Eq.~\eqref{eq:SDP_for_k} using the exclusivity relations of Fig.~\ref{fig:Ex_graph_G6}, we obtain the maximum quantum value
$\beta_{\text{QT}}=\vartheta(\mathcal{G}_6)=\sqrt{5}$.

However, upon computing the dual optimal solution $Z^*$ of the corresponding SDP and constructing the variable matrix $M$ using Eq.~\eqref{eq:M} together with the exclusivity relations of Fig.~\ref{fig:Ex_graph_G6}, it follows that the linear system $MZ^*=0$ admits nontrivial solutions with $M\neq 0$. Consequently, the primal optimal solution of the SDP is not unique. Therefore, this NC inequality does not self-test the corresponding optimal quantum realization as is also shown in Ref.~\cite{KMAAL19}. 

Following our framework, we first construct a composite inequality whose exclusivity graph is obtained as the OR graph product of the exclusivity graphs associated with the individual inequalities, i.e., $\mathcal{G} = C_5 \vee \mathcal{G}_6$, as shown in Fig.~\ref{fig:Ex_graph_parallel_KCBS_G6}. The resulting graph consists of $N = 30$ vertices and $300$ edges. Moreover, we find that $\alpha(\mathcal{G})=4$, while $\vartheta(\mathcal{G})=5$. Using the exclusivity relations of Fig.~\ref{fig:Ex_graph_parallel_KCBS_G6}, we construct the SDP given in Eq.~\eqref{eq:SDP_for_composite}, and numerically obtain a dual optimal solution $Z^*$.

Next, we construct the variable matrix $M$ using Eq.~\eqref{eq:M} together with the exclusivity constraints of $\mathcal{G}=C_5 \vee \mathcal{G}_6$. We then solve the linear system $MZ^*=0$. As before, we reformulate the problem as a homogeneous linear system $Ax=0$ and analyze the corresponding coefficient matrix $A$. In the present case, we take the threshold for identifying zero singular values to $10^{-5}$, corresponding to the numerical precision of the optimization software. After calculating, we find that one singular value is less than the threshold value~\cite{GBS26}, which indicates the coefficient matrix $A$ is numerically rank-deficient. Consequently, the system admits a nontrivial solution $M\neq 0$, implying that the primal optimal solution of the composite SDP is not unique. Therefore, the composite inequality does not self-test the corresponding optimal quantum realization. A numerical package to simulate the scenario is presented in Ref.~\cite{GBS26}.

\end{document}